\documentclass[trackchanges,twocolumn]{aastex701}

\begin{document}

\title{Pulse Morphology and Spectral Study of LPT J1634+44 Using GMRT}

\author[orcid=0009-0002-5290-037X]{Ashish Kumar}
\affiliation{National Centre for Radio Astrophysics, Tata Institute of Fundamental Research, Post Bag 3, Ganeshkhind, Pune-411007, India}
\email[show]{kalyanaastro@gmail.com}  

\author[orcid=0000-0002-0862-6062]{Yogesh Maan} 
\affiliation{National Centre for Radio Astrophysics, Tata Institute of Fundamental Research, Post Bag 3, Ganeshkhind, Pune-411007, India}
\email[show]{ymaan@ncra.tifr.res.in}

\author[orcid=0009-0001-8761-0498]{Ayushi Shrivastava} 
\affiliation{National Centre for Radio Astrophysics, Tata Institute of Fundamental Research, Post Bag 3, Ganeshkhind, Pune-411007, India}
\email{ashrivastava@ncra.tifr.res.in}

\author[orcid=0009-0000-6605-3162]{Banshi Lal} 
\affiliation{National Centre for Radio Astrophysics, Tata Institute of Fundamental Research, Post Bag 3, Ganeshkhind, Pune-411007, India}
\email{blal@ncra.tifr.res.in}

\begin{abstract}
Long-period radio transients (LPTs) are a recently discovered class of Galactic coherent radio emitters characterized by rotation periods of minutes to hours and highly polarized radio pulses. Their nature remains uncertain, with proposed scenarios invoking isolated or binary systems containing neutron stars (NSs) or white dwarfs (WDs). We investigate the nature of CHIME/ILT~J1634+44 through multi-band observations with the Giant Metrewave Radio Telescope (GMRT), examining its pulse morphology, spectral evolution, and timing behavior. The pulses show substantial profile variability, and the subpulse features evolve significantly across observing frequencies. Using our measurements of quasi-periodic microstructures in J1634+44 pulses as well as those of other LPTs available from literature, we show that the scaling relationship between the rotation and microstructure periods is not limited to only neutron stars but extend to white dwarfs as well. The radio spectrum of J1634+44 is highly variable and exhibits a spectral turnover that shifts toward lower frequencies across observing epochs. Interpreting this turnover as free--free absorption, we constrain the physical properties of the intervening absorbing medium. Complementing these emission diagnostics, we also present analysis of pulse arrival times spanning approximately a decade to probe the previously proposed spin--orbit resonance models for this LPT. 
\end{abstract}


\section{Introduction} 
The field of transient radio astronomy is entering an exciting phase, driven by high-resolution, wide-field and sensitive telescopes. Advances in instrumentation and high computing resources have made many discoveries possible. Specifically, better and more efficient time and image-domain searching and real-time data analysis have paved the way for commensal surveys. These advances have opened a discovery space of the long-period radio transients (LPTs), which are observationally manifested by highly energetic and often highly polarized coherent radio pulses, lasting for a few seconds to minutes, and with periodicities ranging from minutes to several hours \citep{Hyman2005, hurley2022}. Due to their wide pulses and intermittent nature, these objects have remained undetected for a long time. The first such type of pulses with a period of $\sim$77\,min were detected towards the Galactic center \citep{Hyman2005, roy10}. A long time after it, the discovery of one more similar object by \cite{hurley2022} reassured the existence of a population of such long-period objects and escalated the search for and further studies of these objects. 
\par
Their extreme brightness temperature, exceeding $10^{12}$\,K, and strongly polarized pulses, reaching almost 100\% in some cases \citep{bvb25}, imply a coherent emission process and suggest an origin from highly magnetized sources or environments. The dispersion measure (DM) values of LPTs are well within the predictions from Galactic Electron Density Models (GEDMs; \citealt{yao2017, oc26}), favoring their galactic origin. These objects often lie below the so-called ``death line'' in the $P-\dot{P}$ phase space (where $P$ and $\dot{P}$ are the spin period and its first time derivative, respectively) challenging conventional pulsar emission models \citep[see reviews by][]{rea2026, caleb2026}. They are proposed to be magnetically powered systems involving highly magnetized compact stars such as neutron stars (NSs) or white dwarfs (WDs), either isolated or in binaries. A few WD binary systems emit periodic radio emission with lower polarization and orders of magnitude less luminosity than LPTs \citep{marsh2016,buckley2018,pelisoli2023}. Similarly, magnetars also show periodic and polarized radio emission, but their on-average shorter periods do not exactly fit in this category \citep{kaspi2017}. However, rare magnetars with very long periods \citep[e.g., 1E161348$-$5055 has a rotation period of 6.7\,hours,][]{dai16}, and the more recently discovered PSR~J0901$-$4046 with 76\,s period, keep the NS as a possible contestant for LPTs \citep{Caleb22}. 
\par
To date, nineteen LPTs have been discovered, of which four sources, ILT J1101+5521 \citep{ruiter2025}, GLEAM-X J0704-37 \citep{hurley2024,rodriguez2025}, ASKAP J174508.9-505149 \citep{rose2026}, and ASKAP J1448-68 \citep{mo26}, have been identified as WD binaries through observations of their optical counterparts. These objects exhibit strong pulse-to-pulse variability in their spectro-temporal profiles and polarization behavior, and also show broad spectral properties (spectral index varies from 0.4 to -13.5; e.g., \citealt{rea2026}). All these observed properties do not fit in a currently known single source class of coherent emitters, and hence, there is no unified model for the LPT progenitors. 
\par
CHIME/ILT J1634+44 (hereafter J1634+44) has been independently discovered in LOFAR archival observations during 2015--2019 \citep{bvb25}, and in CHIME observations during 2020--2024 \citep{dong25_J1634}. Using VLA observations, \cite{dong25_J1634} localized the source with arcsecond precision to (RA, Dec) = (16$^{\mathrm{h}}$34$^{\mathrm{m}}$29.96$^{\mathrm{s}}$, +44$^{\circ}$50$^{\prime}$13.5$^{\prime\prime}$). The DM value of J1634+44 is measured to be 25.0(2)\,pc\,cm$^{-3}$, corresponding to a distance of 1.4--3\,kpc. The source shows strong pulse-to-pulse variability, with flux densities ranging from a few mJy to 12\,Jy. Interestingly, pulse timing analysis of J1634+44 shows two periodicities, 841\,s and 4206\,s, with significant spin-up, which does not fit the isolated dipole radiation model and suggests a binary system interpretation \citep{bvb25,dong25_J1634}. Its pulses exhibit quasi-periodic microstructures \citep{dong25_J1634}. The emission from J1634+44 is highly polarized with significant variability; \cite{bvb25} have reported nearly 100\% linear and as well as 100\% circular polarization in different pulses, whereas \cite{dong25_J1634} have detected only circular polarization, with no significant linear polarization component. Based on optical counterpart detection and photometric analysis, \cite{bvb25} have proposed that J1634+44 is a WD binary system, with another WD, NS, or Ultra-cool dwarf (UCD) as a companion. On the contrary, \cite{dong25_J1634} have proposed J1634+44 to have a NS-like progenitor.
\par
Given the diverse range of properties exhibited by J1634+44, it is an interesting object to follow up. Through our observations covering wide frequency range using the Giant Metrewave Radio Telescope (GMRT; \citealt{swarup1991, gak17}), we present here our probes of its various emission characteristics. We investigate pulse-to-pulse spectro-temporal variations, including the quasi-periodic microstructures and spectral behavior. Rest of the paper is structured as follows. In Section~\ref{sec:obs}, we have summarized the observation and the search procedure. Section~\ref{sec:analysis} describes the data analysis and our findings for J1634+44. The discussion and the conclusion are presented in Sections~\ref{sec:discussion} and \ref{sec:conclusion}, respectively.

\section{Observations and Search Procedure}\label{sec:obs}
We observed J1634+44 using the GMRT for a total of 20 hours between 20 September 2025 and 1 July 2026. Of these, 12\,hours were utilized under the Director’s Discretionary Time (project code: DDTC451) and 8\,hours under a regular proposal (project code: C50\_120). The GMRT consists of 30 parabolic dish antennas, each of 45\,m diameter, with a maximum baseline of 25\,km \citep{swarup1991, gak17}. The observations were carried out across six epochs. Two high sensitivity observations utilized single-band (band 3, 300-500\,MHz), and the rest were simultaneous in bands 3, 4 (550-750\,MHz), and 5 (1260-1460\,MHz) using the subarray capability of the GMRT. These observations were performed in phased array mode, using $\sim20$ antennas in single-band and $\sim8$ antennas per subarray in multi-band mode. The farthest outer-arm antennas were excluded to maintain phase coherence during beam-forming. All the observations typically utilized 2048 channels across 200 MHz bandwidth (in individual bands, in case of multi-frequency observations) and a sampling time of 163.84\,$\mu$s. In the two single-band observations, a higher time resolution of 81.92\,$\mu$s was employed.

\subsection{Data Reduction and Single Pulse Detection}
We use \texttt{RFIClean}\footnote{\url{https://github.com/ymaan4/RFIClean}} \citep{MvLV21} to excise radio frequency interference (RFI) and convert raw beamformed GMRT data into SIGPROC filterbank format. RFIClean excises periodic, narrowband as well as broadband spiky RFI. The resulting data are subsequently corrected for bandpass shape using \texttt{digifil} from \texttt{DSPSR} package \citep{vsb11}. Further data processing is carried out utilizing \texttt{PRESTO} software suite \citep{ransom11}. A second round of RFI excision is applied using \texttt{rfifind} from \texttt{PRESTO}. The cleaned data are then initially de-dispersed with the known DM of 25\,pc\,cm$^{-3}$ \citep{dong25_J1634} using \texttt{prepdata}. Using a narrow component in one of our first detected pulses, we estimated the DM to be 25.09\,pc\,cm$^{-3}$ using \texttt{pdmp} from \texttt{PSRCHIVE} \citep{hoton2004}. We then used this updated DM value in all of our searches. We searched for pulses in the dedispersed time series using \texttt{single\_pulse\_search.py}, which first detrends the time series to remove slow baseline variations and then convolves the data with boxcar filters of different widths and identifies candidate pulses based on their signal-to-noise ratio (S/N). Since LPT pulses are typically several seconds wide, depending on the detrending timescale, some of broad features or the whole pulses might get subtracted as baseline variations. Furthermore, the statistics for different boxcar widths are extrapolated from that of the native resolution time series assuming a gaussian noise. However, this assumption rarely holds for real data and, especially for large boxcar widths, the extrapolated statistics deviates significantly from the actual. To circumvent these issues and enhance sensitivity to broad pulses, we first prepare downsampled versions of the filterbank data with the original resolution using factors of 8, 16, and 32, and then search the downsampled filterbank files separately using the same procedure. For simultaneous multi-band observations, we have searched for pulses in each band separately, and have also taken a deeper look at the corresponding data in other bands whenever a pulse is detected in one of the bands. Surprisingly, in our simultaneous multi-band observations, pulses have not been detected at band 5, but are detected at other bands. No detection could be made at band 5 despite manual examinations and searching for pulses by convolving the expected band 5 pulse regions with boxcars of different widths.
\par
\begin{figure*}
    \centering
    \includegraphics[width=0.497\linewidth]{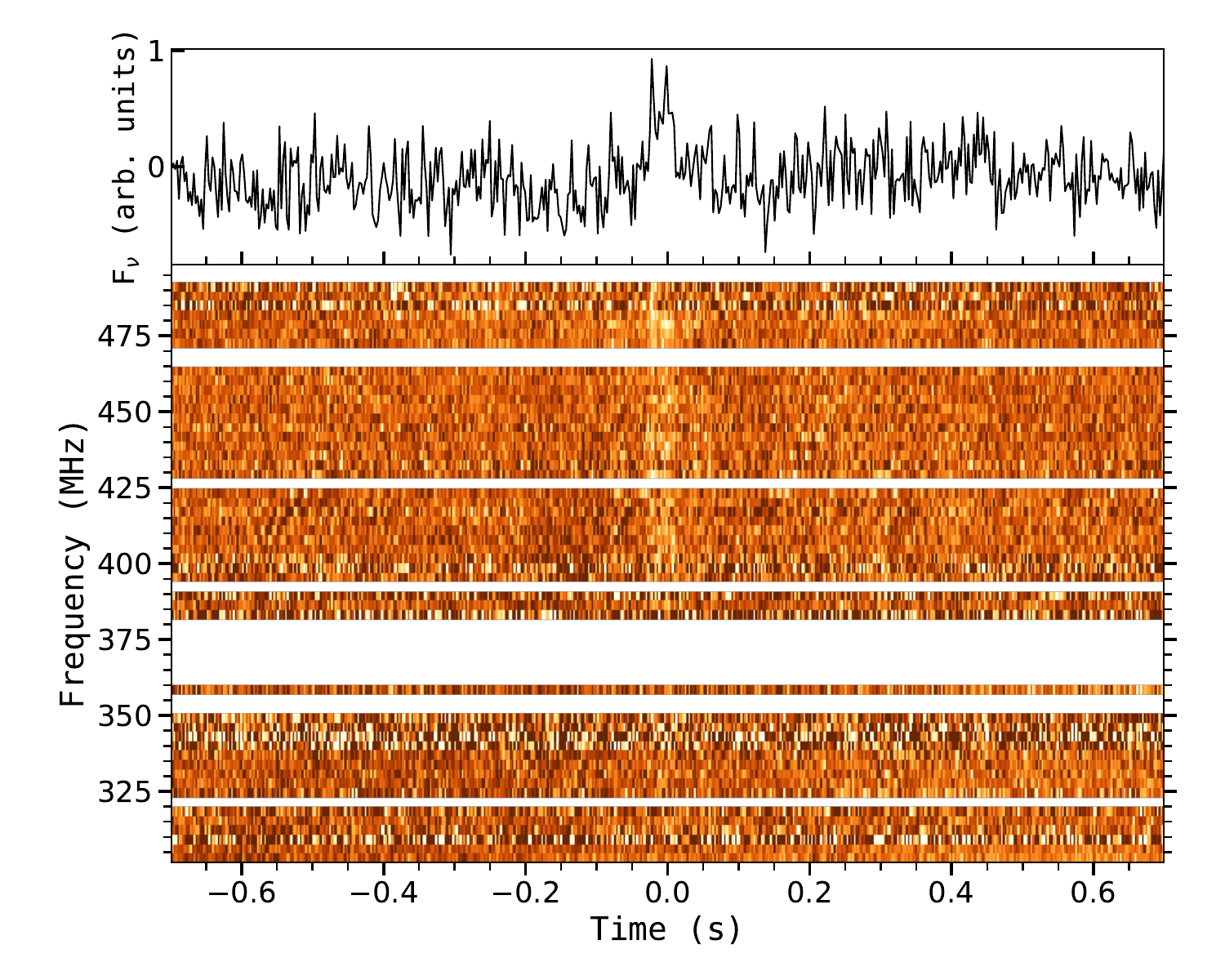}
    \includegraphics[width=0.497\linewidth]{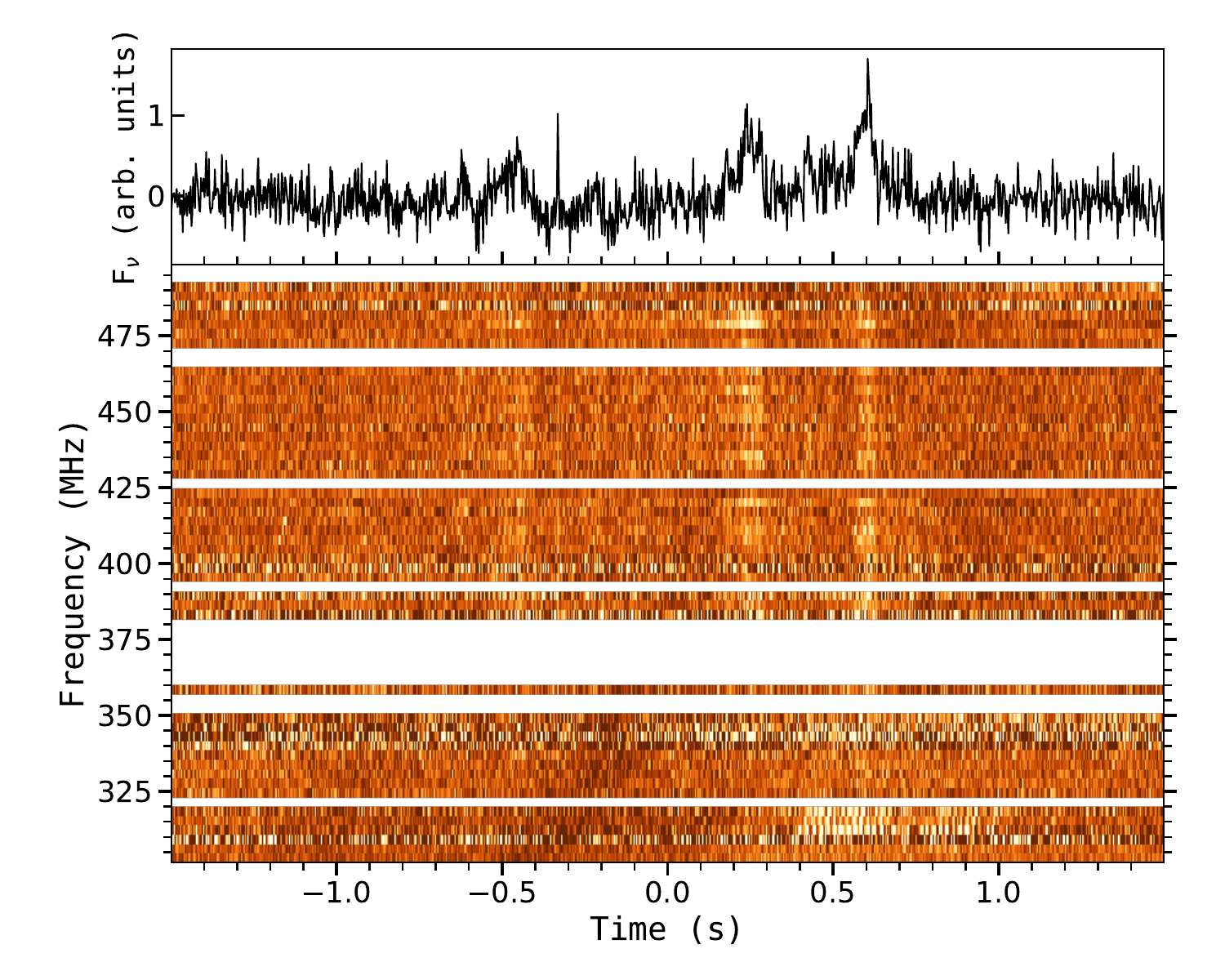}
    \includegraphics[width=0.497\linewidth]{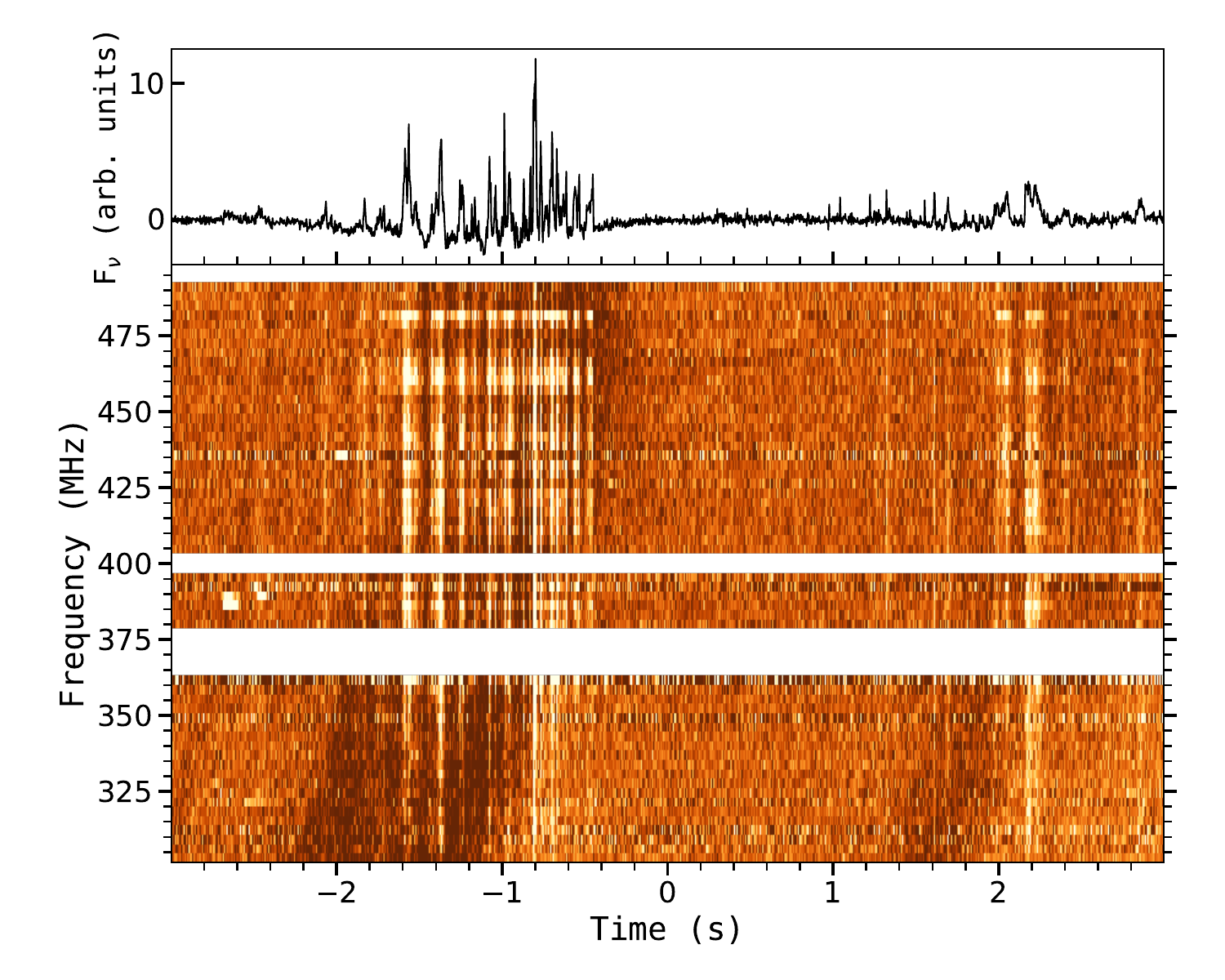}
    \includegraphics[width=0.497\linewidth]{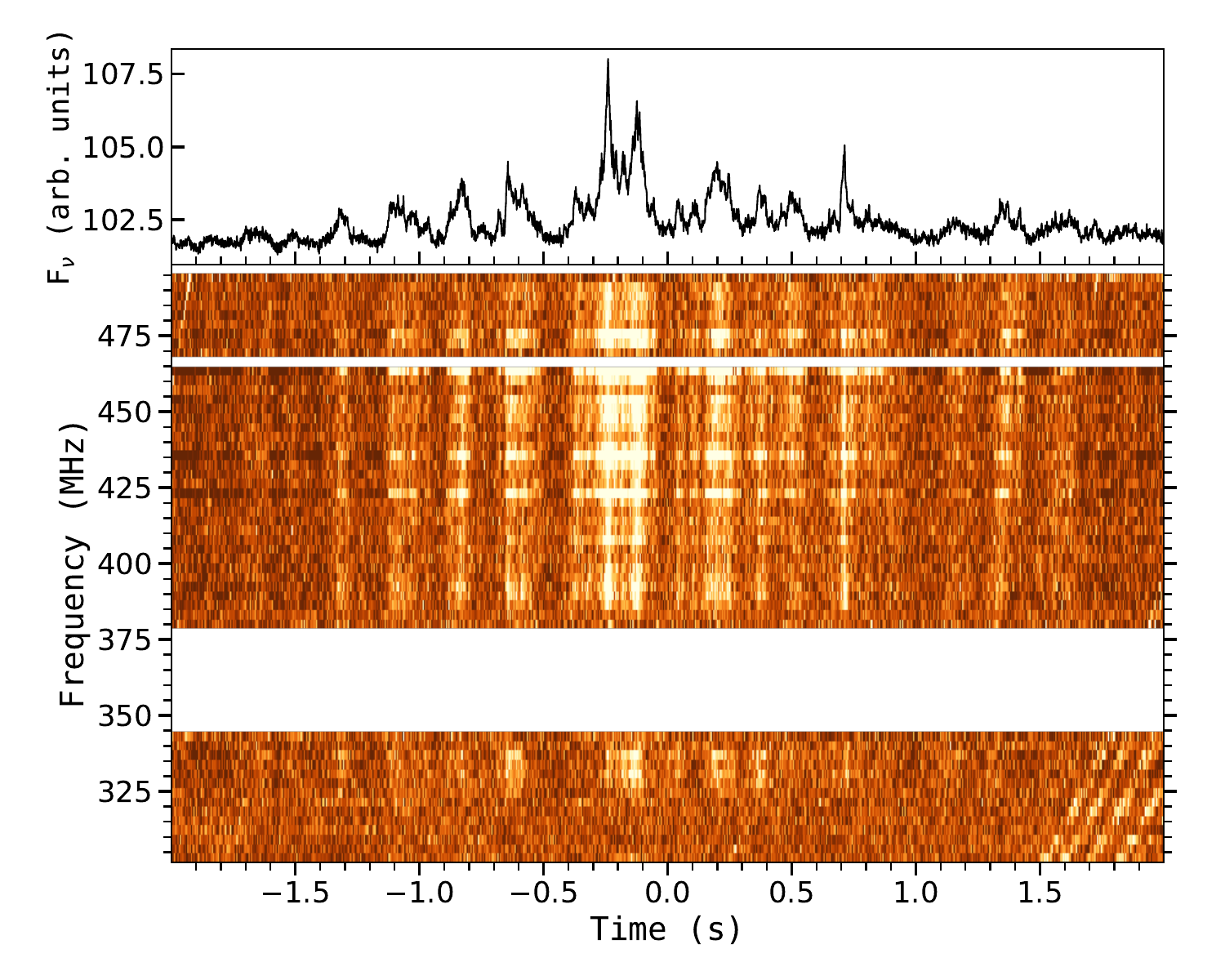}
    \caption{The figure highlights a sample of detected pulses from J1634+44 in our GMRT observations. The gallery of pulses shows different pulse morphologies, belonging to C1, C2, C3, and C4, from top left to bottom right, respectively (see Section \ref{sec:spm}). In each plot, the bottom panel shows the dynamic spectrum of the de-dispersed pulse, and the top panel shows the uncalibrated, frequency averaged de-dispersed pulse.}
    \label{fig:pulse_gallery}
\end{figure*}

For all the detected pulses, we estimate DM using \texttt{DM\_phase}\footnote{\url{https://github.com/danielemichilli/DM_phase}} package \citep{smp19}, which determines the optimal DM by maximizing the coherent power and is well suited for pulses with complex morphology. These DM estimates are presented in Table~\ref{tab:bp}. The earlier mentioned \texttt{pdmp} estimate of 25.09(1)\,pc\,cm$^{-3}$ is consistent with the average value of DM estimated using \texttt{DM\_phase}. For the rest of the analysis presented in this work, all the detected pulses have been de-dispersed using a DM value of 25.09\,pc\,cm$^{-3}$, and examples of detected de-dispersed pulses are shown in Figure~\ref{fig:pulse_gallery}.

\section{Data Analysis and Results}\label{sec:analysis}
We have detected a total of 10 pulses in our observing campaign. Mostly two pulses have been detected in each DDT observation, but no pulse has been detected in the last two observations. In two simultaneous multi-band DDT observations, four pulses are detected in bands 3 and 4, but none in band 5. The remaining six pulses are detected in two single-band observations. Using these pulses, we have conducted a detailed investigation of both the spectral properties and spectro-temporal pulse morphology of J1634+44. Further, we present refined timing solution for this object and provide insights about the spin-orbit resonance model \citep{bvb25}. The details of these analyses are presented in the following subsections.

\subsection{Flux Density and Spectral Index}\label{sec:flux_d_si}
\begin{figure*}
    \centering
    \includegraphics[width=0.497\linewidth]{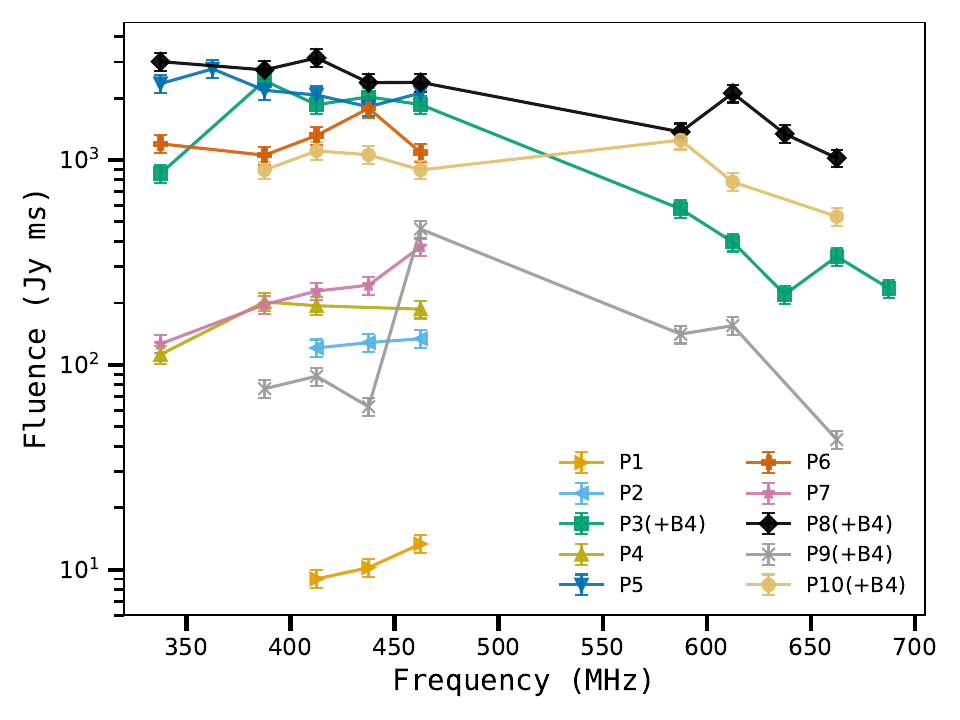}
    \includegraphics[width=0.497\linewidth]{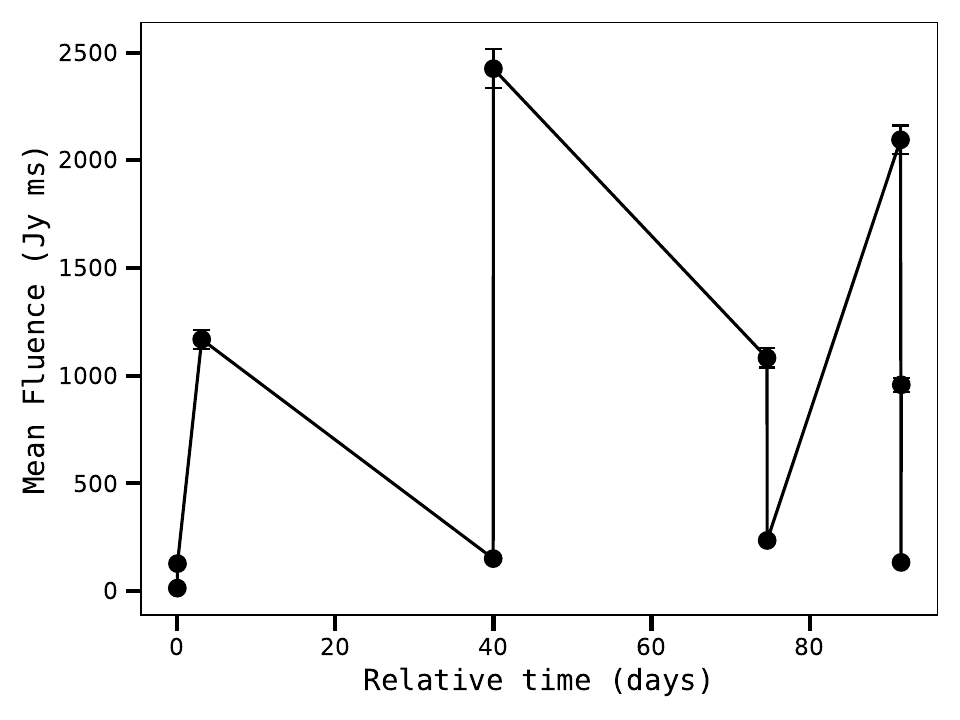}
    \caption{The figure shows fluence variations across different epochs. Left: Data points represent fluence estimates at the respective subbands. Four pulses (3, 8, 9, and 10) are detected simultaneously at bands 3 and 4, thereby spanning a wider frequency range. Right: The average fluence across different epochs with associated propagated measurement uncertainty. The x-axis represents relative time with respect to the first pulse detected in our observing campaign.}
    \label{fig:flu_vary}
\end{figure*}

We estimate flux density using the radiometer equation \citep{lk04}, which requires tye system equivalent flux density (SEFD\footnote{SEFD = T$_{\rm sys}$/G, where T$_{\rm sys}$ and G are the system temperature and the antenna gain, respectively.}). We model SEFD as a function of frequency using a polynomial whose coefficients are provided by the observatory\footnote{\url{www.gmrt.ncra.tifr.res.in/gtac/etc/etc/etc_help.pdf}}. The estimated SEFD is expected to carry an uncertainty of about 5–10\%; we conservatively adopt a 10\% uncertainty in the SEFD, which propagates to both the flux density and the fluence estimates. The frequency-dependent background sky temperature ($\rm T_{sky}$) is obtained using \texttt{skytempy}\footnote{\url{https://libraries.io/pypi/skytempy}}, which uses the 408\,MHz Haslam sky map \citep{hss8,rbb15}. The flux density calibration is performed on subbanded data, by dividing each observing band into eight subbands. Further, to minimize the bandshape effects, 5\% of the total channels at the band edges are excised. From the calibrated data, the fluence is calculated by integrating the flux density across the pulse after subtracting the baseline. To estimate baseline variations, we select two off-pulse regions on either side of the pulse and compute noise statistics (median and RMS) at ten sub-intervals in each region. These noise statistics are then interpolated using a second-order polynomial fit across the pulse bins, and the resulting temporal baseline is subtracted from them. The estimated fluence varies significantly over epochs, and there is no apparent trend in them (Fig.~\ref{fig:flu_vary}). Further, we estimate the peak flux density, which represents the maximum value of the pulse time series convolved with the optimal boxcar of unit height. The optimal boxcar width itself is determined by convolving the pulse with boxcar filters of varying widths and selecting the width that maximizes S/N. Following \cite{dong25_J1634}, we also report the effective pulse width, defined as the ratio of the fluence to the peak flux density. It is equivalent to the width of a top-hat pulse with a height of peak flux density and containing the total pulse energy. 
\par
The wide fractional bandwidths of our observations allow to estimate the in-band spectral index. The spectral index for each detected pulse is estimated by fitting a power law, $S_\nu = A(\nu / \nu_{\rm ref})^\alpha$, where $A$, $\nu_{\rm ref}$, and $\alpha$ are the amplitude, reference frequency (fixed at 500\,MHz), and the spectral index, respectively, to fluence values estimated at individual subbands. The noise-statistics in the edge subbands are generally severely affected by the band shape, and therefore, these are excluded from the spectral fitting. We have also excluded the subbands wherein more than 60\% of channels within the subband were RFI-contaminated. The spectral fitting to the fluence values estimated from single band data mostly follows a power law. For majority of pulses detected at band 3, the spectrum is inverted ($\alpha>0$), and it becomes steep for pulses detected at band 4 (see Table~\ref{tab:bp}), suggesting a spectral turnover. Therefore, to estimate the turnover frequency, we fit a log-parabola to the spectrum. We have used the following mathematical form of a log-parabola:
\begin{equation}
    \log_{10}S_\nu = \beta \times (\log_{10}\nu - \log_{10}{\nu_p})^2 + C_0,
    \label{eq:log_para}
\end{equation}
where $S_\nu$ is the fluence, $\beta$ is the curvature of the parabola, $\nu$ and $\nu_p$ are the center frequency of the subband and the turnover frequency, respectively, and $C_0$ is a constant. 
\par
\begin{figure*}
    \centering
    \includegraphics[width=0.497\linewidth]{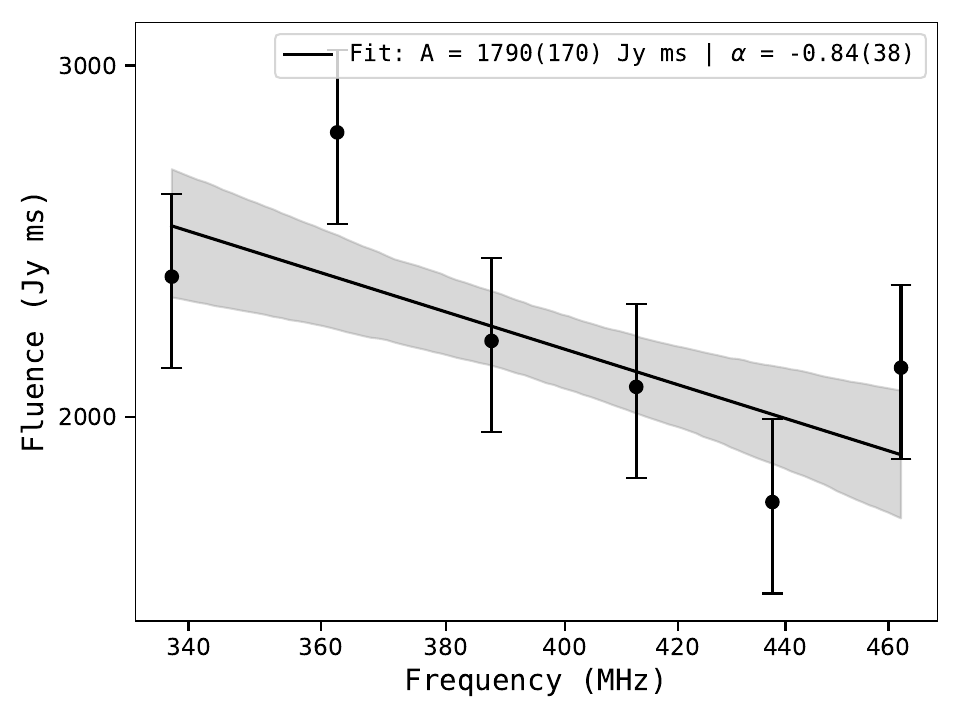}
    \includegraphics[width=0.497\linewidth]{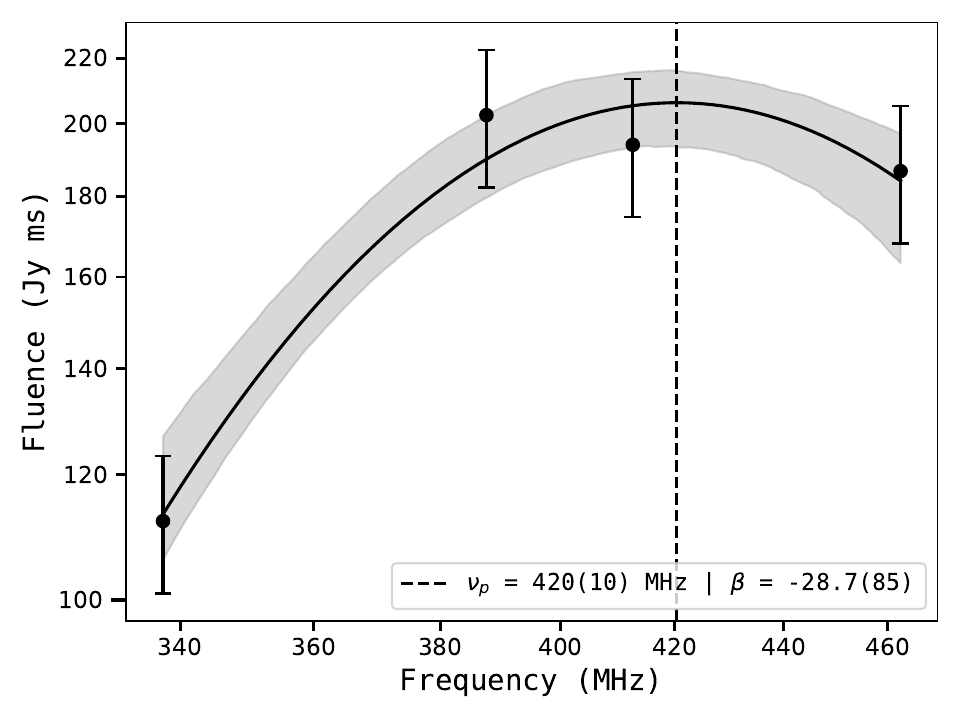}
    \includegraphics[width=0.497\linewidth]{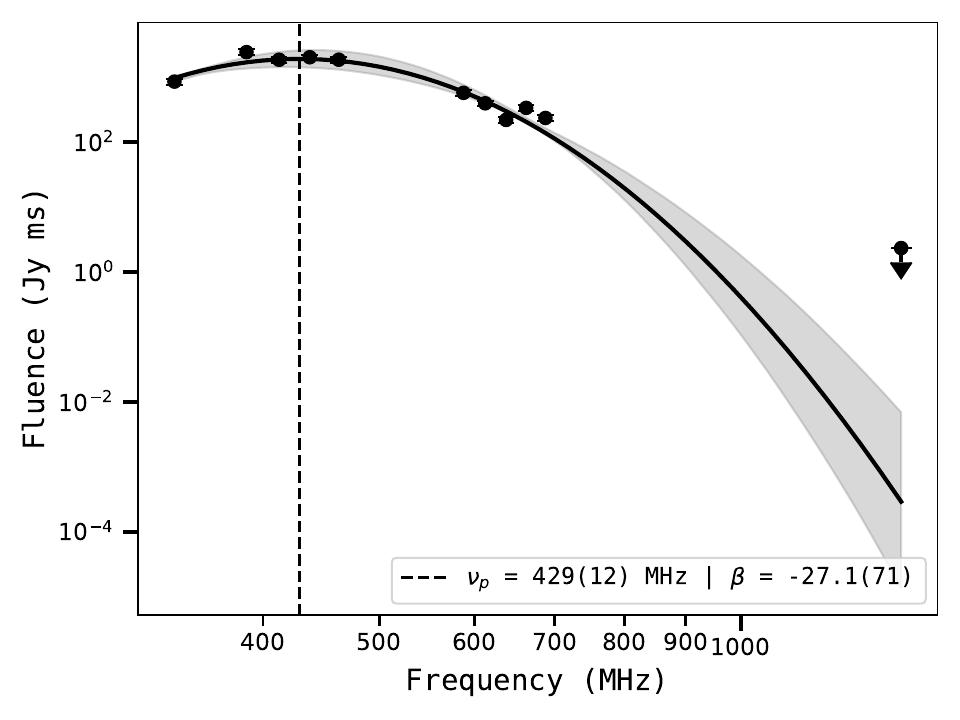}
    \includegraphics[width=0.497\linewidth]{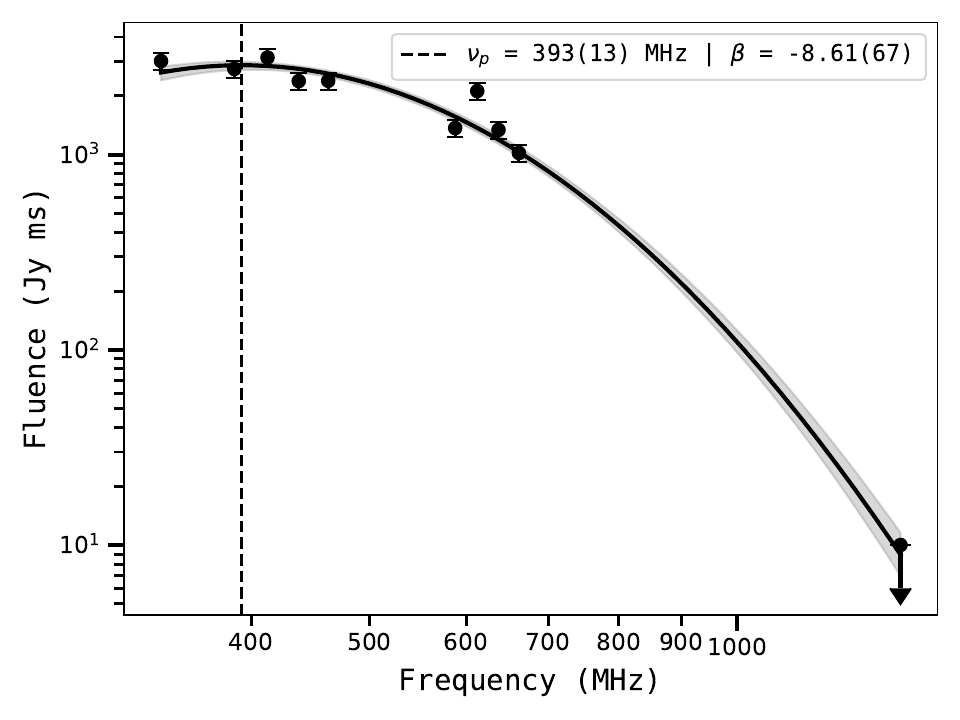}
    \caption{The figure presents examples of spectral fitting to fluence values estimated from single-band and multi-band observations. The top left plot shows an example of a power-law fitting using a single band observation. The top right plot shows an example of a log-parabola fit to the single-band fluence values. The bottom row panels show spectral fitting using simultaneous multi-band data at different epochs, including the upper limits from band 5 (for details, see Section~\ref{sec:flux_d_si}). In all the panels, the continuous line and the shaded region around it represent the best-fit model and the associated 1\,$\sigma$ uncertainty, respectively. The fitted parameters are shown in the plot legend, and the vertical dashed line marks the turnover frequency.}
    \label{fig:spectral_fit_log}
\end{figure*}
\begin{figure*}
    \centering
    \includegraphics[width=0.6\linewidth]{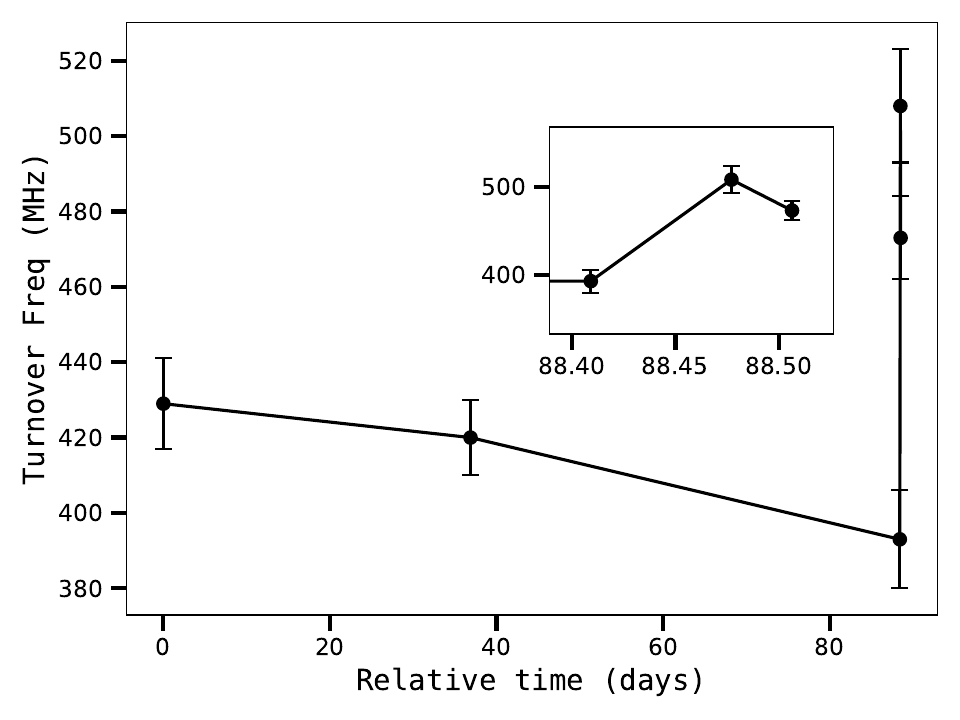}
    \caption{The plot shows turnover frequency evolution over epochs, and the inset shows a zoomed-in view of the last three turnover frequencies. The x-axis represents relative time with respect to the first data point in the plot, which is at MJD~60941.663442.}
    \label{fig:nu_p}
\end{figure*}

At two single band epochs, the power law does not fit the data well. For these epochs, we found that the log-parabola better fits the fluence values (Fig.~\ref{fig:spectral_fit_log}). For multi-band data spectral fitting, we have also included the upper limit on fluence from band 5, following an approach similar to that used by \citet{maan2026}. The spectral parameters are then estimated using a maximum likelihood estimation (MLE) approach that simultaneously accounts for both detected fluence measurements and the upper limit. For measured data points, the likelihood is defined by a Gaussian probability density function, which measures the proximity of the data to the model. For non-detections, the fitting utilizes the cumulative distribution function to constrain the model curve to remain statistically consistent with the observed upper limit (i.e., below). We have considered a conservative 10\% uncertainty to the upper limit. Fig.~\ref{fig:spectral_fit_log} shows examples of the multi-band spectral fitting; the fitted spectral turnover frequencies range in (393, 508)\,MHz, with a mean value of 445\,MHz (see Fig.~\ref{fig:nu_p}). The uncertainties in the fitted parameters from both the power-law and log-parabola fits are estimated using a Monte Carlo simulation. We have generated 1000 synthetic parameter sets by sampling from a multivariate normal distribution defined by the best-fit parameters and their associated covariance matrix. The 1$\sigma$ confidence interval is then derived from the 16$^{\rm th}$ and 84$^{\rm th}$ percentiles of the distribution of resulting model realizations.

\begin{deluxetable*}{rllllllllll}
\digitalasset
\tablewidth{0pt}
\tablecaption{The estimated parameters for all detected pulses. The columns represent the pulse number, the topocentric pulse ToA, GMRT observing band, estimated DM, fluence, peak flux density, effective pulse width, the power law spectral index, width and period of quasi-periodic microstructures, and pulse category. 
\label{tab:bp}}
\tablehead{
No. & \colhead{ToA} & Band & \colhead{DM} & \colhead{Fluence} & \colhead{$S_{\rm Peak}$} & \colhead{$W_{\rm eff}$} & \colhead{$\alpha$} & \colhead{$W_\mu$} & \colhead{$P_\mu$} & Category\\
 & (MJD) & & \colhead{(pc\,cm$^{-3}$)} & \colhead{(Jy\,ms)} & \colhead{(mJy)} & \colhead{(ms)} & & \colhead{(ms)} & \colhead{(ms)} &
}
\startdata
P1 & 60938.5768595(5)$^{1}$ & 3 & -- & 13(1)& 109(11) & 122(17) & 3(1) &--&--& C1\\
P2 & 60938.59629(1) & 3 & 25.51(86) & 127(13)   & 94.8(95) & 1340(190) & 1(1)                    &--&--& C2\\
P3 & 60941.66344(2) & 3 & 24.98(02) & 1850(180) & 919(92)  & 2010(280) & 2.5(4)$^{3}$ &--&--& C3\\
P3 & 60941.66344(2) & 4 & 24.99(19) & 353(35)   & 233(23)  & 1510(210) & -5(1)                   &--&--& C3$^{2}$\\
P4 & 60978.53683(1) & 3 & 24.98(08) & 150(15)   & 160(16)  & 940(130)  & 1.6(4)$^{3}$ &--&--& C4\\
P5 & 60978.56614(2) & 3 & 25.06(01) & 2430(240) & 856(86)  & 2830(400) & -0.8(4) & 163.2 & 321.4 & C4\\
P6 & 61013.17975(1) & 3 & 24.99(14) & 1080(110) & 348(35)  & 3110(440) & 0.2(4)                  &--&--& C4\\
P7 & 61013.19913(2) & 3 & 25.32(21) & 235(23)   & 164(16)  & 1430(200) & 3.1(4) & 74.7 & 175 & C4\\
P8 & 61030.07249(2) & 3 & 25.11(03) & 2460(250) & 1530(150)& 1610(230) & -0.8(4)& 90.9 & 177.4 & C4\\
P8 & 61030.07248(2) & 4 & 25.27(01) & 1460(150) & 621(62)  & 2350(330) & -3(1)$^{3}$  &--&--& C4\\
P9 & 61030.140542(8)& 3 & 24.61(84) & 143(14)   & 198(20)  & 720(100)  & 0(1)$^{3}$   &--&--& C3\\
P9 & 61030.140539(8)& 4 & -- & 113(11)  & 188(19)   & 601(85)& -10(1)         &--&--& C3\\
P10& 61030.16985(2) & 3 & 24.83(01) & 1010(100) & 481(48)  & 2100(300) & -0.01(71)$^{3}$ & 112.7 & 218.8 & C4\\
P10& 61030.16985(6) & 4 & 24.91(15) & 852(85) & 456(46) & 1870(260) & -7(1)                     &--&--& C4\\
\enddata
\tablecomments{
$^{1}$ This ToA is estimated after smoothing the pulse profile with a 6\,ms Gaussian kernel (see Section \ref{sec:timing}).\\
$^{2}$ It is heavily affected by RFI\\
$^{3}$ Log-parabola fits the data better than power law.
}
\end{deluxetable*}


\subsection{Single Pulse Morphology}\label{sec:spm}
Pulses from transient sources generally exhibit rich substructures that encode valuable information about the underlying emission mechanism, the local magneto-ionic environment, and the propagation effects introduced by the intervening medium. Pulses from J1634+44 also contain substructures on a few tens (or finer) to several hundred millisecond timescales, which vary from pulse to pulse. Motivated by the pulse classification approach used by \citet{Caleb22} and \citet{Dong24_thesis}, we have classified the pulses into the following four categories: C1: single peak pulse, C2: triple-peak pulse, C3: pulse with two emission episodes, and C4: broad multi-peak pulse with the emission appearing to cease or reach close to the baseline in between the peaks (Fig.~\ref{fig:pulse_gallery}). In our sample of detected pulses, one pulse belongs to each of the C1 and C2 categories, whereas two and six pulses belong to the C3 and C4 categories, respectively. Here, the pulses simultaneously detected in bands 3 and 4 are counted only once. Four pulses belonging to the C4 category exhibit clear quasi-periodic microstructures, which are described in Section~\ref{sec:qp}. 
\par
\begin{figure*}
    \centering
    \includegraphics[width=0.497\linewidth]{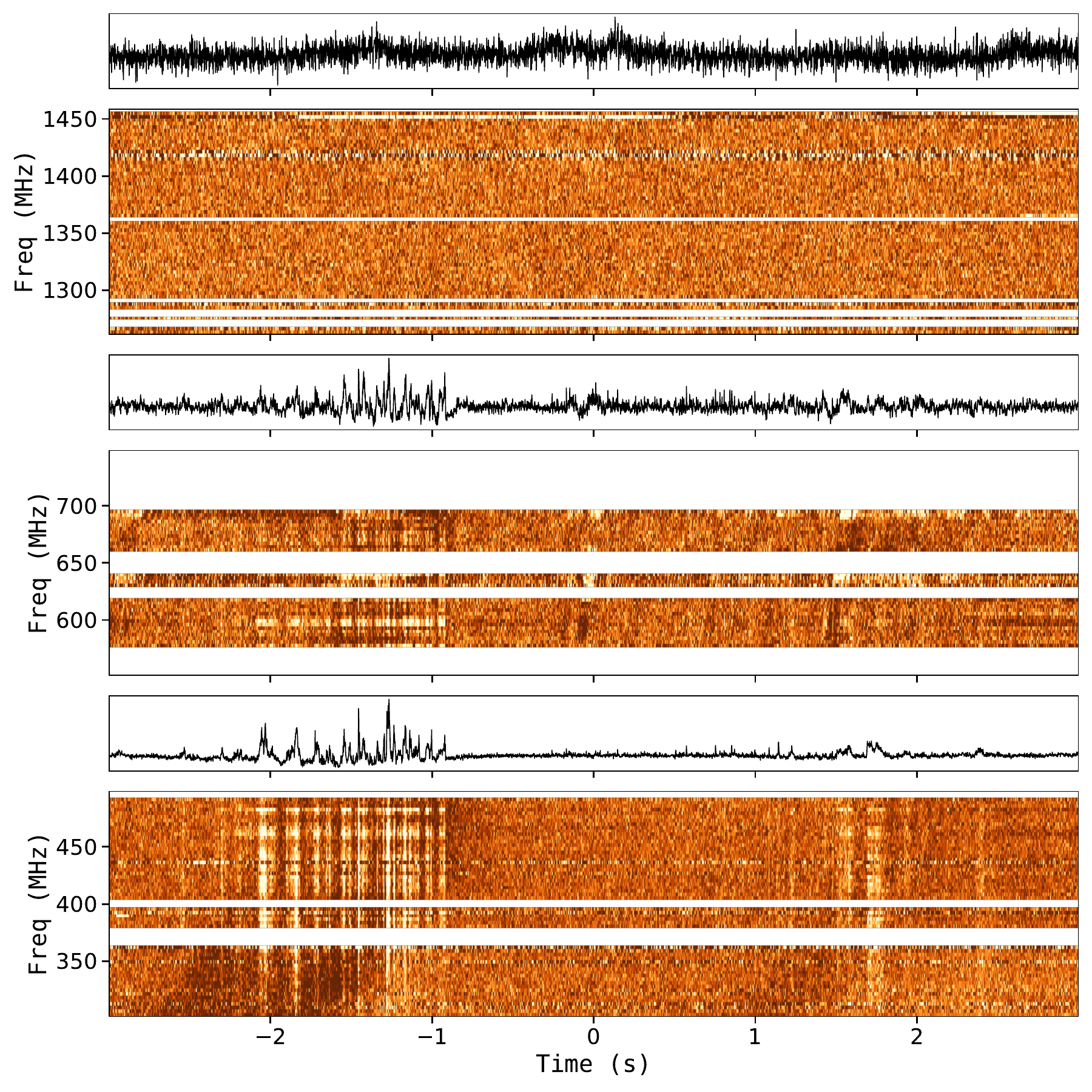}
    \includegraphics[width=0.497\linewidth]{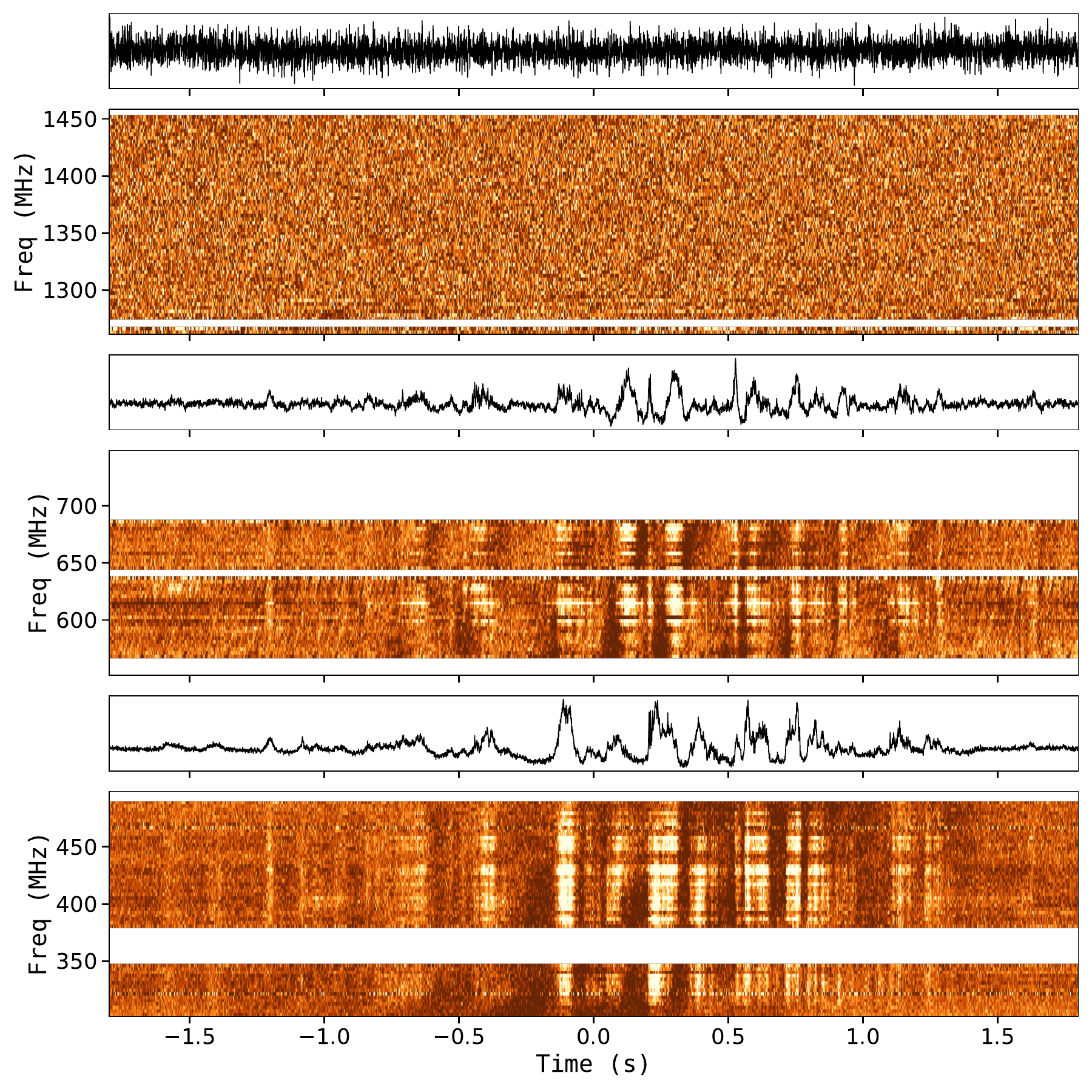}
    \caption{The figure illustrates simultaneous multi-band detection of pulses at two epochs. The x-axis represents the time from the start of the observation. The stacked pulses, from bottom to top, belong to bands 3, 4, and 5, respectively. The pulse is not detected at band 5. These pulses are temporally aligned with respect to the highest frequency of band 5 (1460\,MHz). The masked regions in time-series and dynamic spectra are due to the RFI flagging.}
    \label{fig:pulses_freq_evo}
\end{figure*}

The subpulses exhibit significant temporal variability, even just over a couple of spin periods. The pulses shown in the top panel of Fig.~\ref{fig:pulse_gallery} belong to the same observation, and just after two rotations, the pulse has evolved from a single component to a three-component pulse. In addition to temporal variations, we study frequency-dependent pulse evolution of four pulses, which are detected simultaneously in bands 3 and 4, but not detected in band 5. Two of such pulses are shown in Fig.~\ref{fig:pulses_freq_evo}. These pulses are temporally aligned with the highest frequency of band 5, accounting for the dispersive delay introduced by the ionized interstellar medium (ISM) and any delay in data recording. Notably, these pulses reveal that the subpulse features evolve with frequency. Some of the subpulses diminish or split across frequency bands. The relative flux density of the subpulses also varies across different frequency bands, i.e., a comparatively weaker subpulse in band 3 becomes comparatively stronger in band 4, and vice versa. Importantly, in one pulse, we have noticed a sharp cutoff to the emission profile (bottom left plot in Fig.~\ref{fig:pulse_gallery}). Overall, J1634+44 exhibits an intriguing evolution in its spectro-temporal pulse profiles, whose possible physical interpretations are discussed in Section~\ref{sec:disc_pulse_mor}.

\subsubsection{Quasi-periodic Microstructures}\label{sec:qp}
\begin{figure*}
    \centering
    \includegraphics[height=0.45\linewidth]{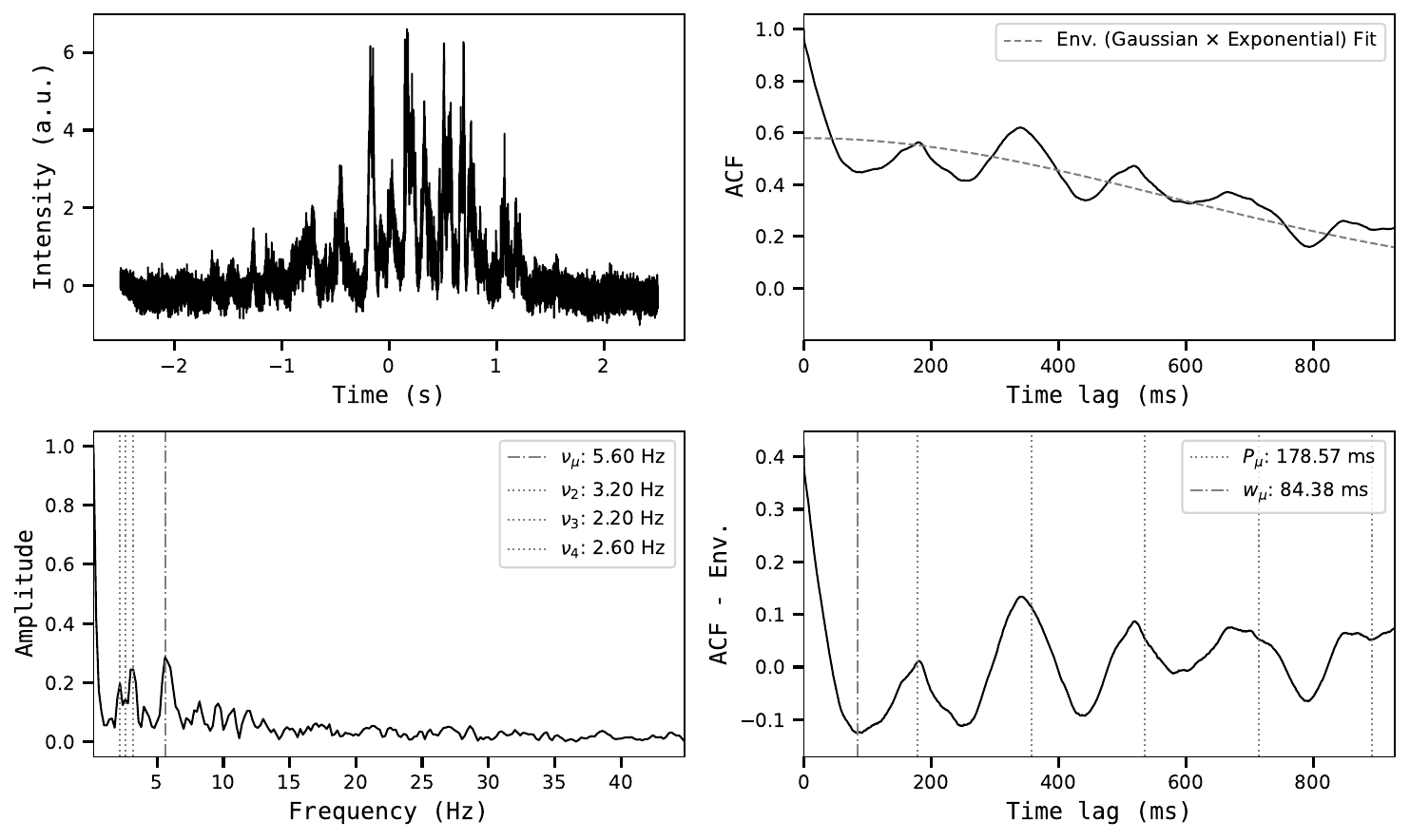}
    \includegraphics[height=0.65\linewidth]{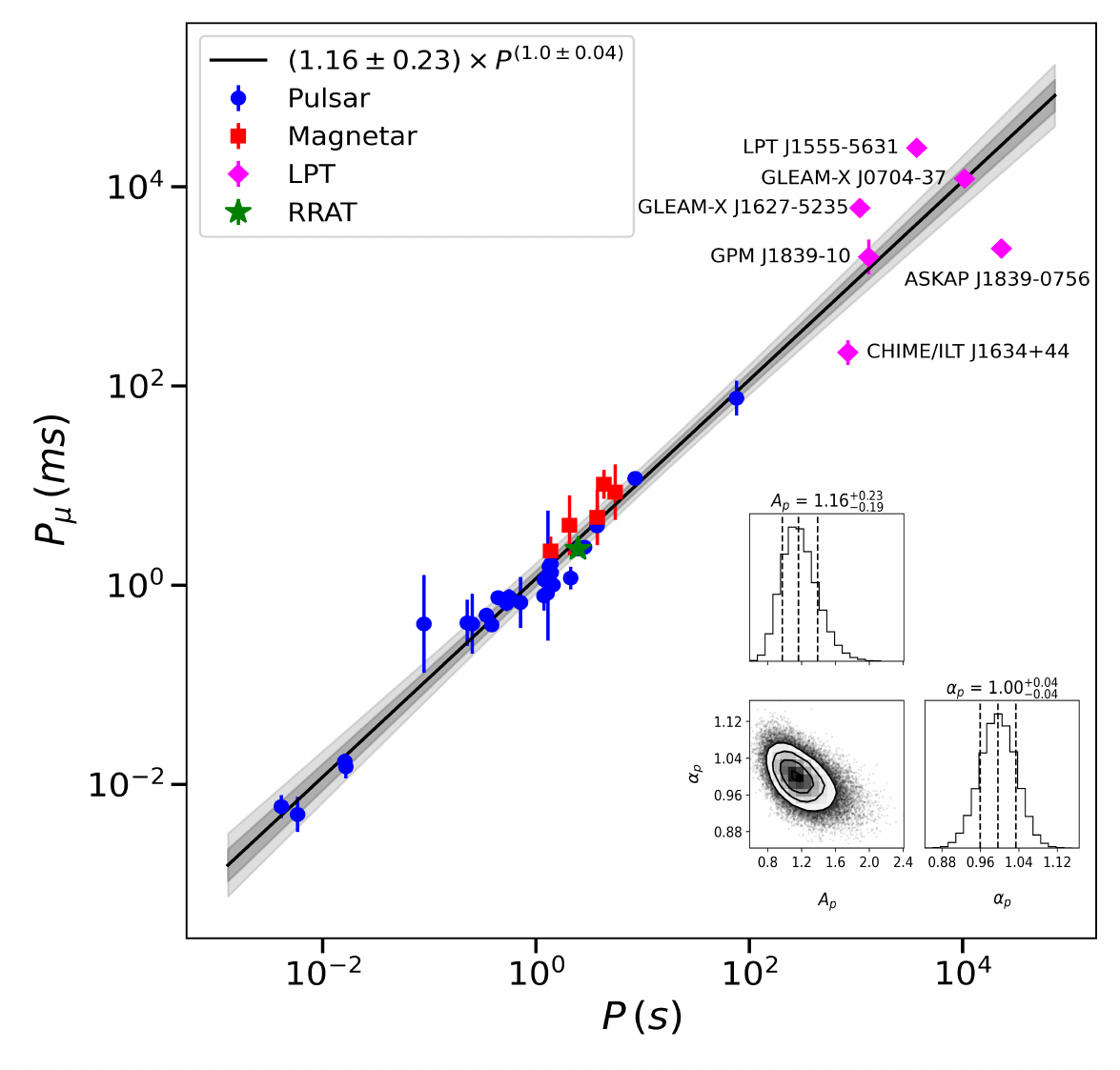}
    \caption{Top: Pulse microstructure analysis of J1634$+$44. From top left to bottom right, the different panels show a) pulse time series where the intensity is in arbitrary unit; b) ACF of the pulse time series, overlaid with a dashed line, representing the best fitted Gaussian multiplied one-sided exponential function fitted to the ACF; c) the Fourier transform of the pulse time series overlaid with vertical lines indicating significant peaks (he frequency corresponding to the most significant quasi-periodicity is denoted by a dashed-dotted vertical line); d) ACF residuals after subtracting the best fitted envelop function, where the dotted vertical lines mark the quasi-periods (P$_\mu$, 2P$_\mu$, and so on) and dashed-dotted line marks the width of the quasi-periodic oscillations. Bottom: Empirical relationship between $P$ and $P_\mu$ for pulsars, magnetars, RRATs, and LPTs. The solid line represents the best-fitting power-law relation, and the shaded regions indicate the $1\sigma$ and $2\sigma$ uncertainty intervals. The inset shows the corner plot of the fitted parameters, where $A_p$ is the scaling factor and $\alpha_p$ is the power-law index.}
    \label{fig:microst}
\end{figure*}

High temporal resolution observations of pulsars and transient sources have revealed fine sub-pulse features and quasi-periodic intensity modulations with characteristic timescales ranging from microseconds to sub-seconds, commonly referred to as microstructures or substructures. The common methods to identify the quasi-periodic structures include autocorrelation, derivative of the autocorrelation function, and Fourier transform \citep{cordes1979, kramer02, amarnath26}. As shown for one of our pulses (among four) in Fig.~\ref{fig:microst}, the auto-correlation function (ACF) clearly reveals quasi-periodic behavior in the underlying substructures. The substructure period is determined by identifying the significant peaks in the Fourier spectrum of the time series. The peaks and troughs in the ACF becomes more apparent after a slowly varying envelop is removed. In Fig.~\ref{fig:microst}, a best fitted Gaussian multiplied by one-sided exponential function is subtracted from ACF, which is shown as ACF residuals. Further, the substructure width is estimated as the time lag corresponding to the first minimum in ACF residuals. Some care needs to be taken while interpreting the width of the quasi-periodic structures estimated from the ACF, particularly when the estimated width is half of the substructure period. In such a case, if the separation between the substructures is less than their widths, then the ACF does not carry any meaningful information about the actual width \citep{Mitra_2015}. 
\par
In four pulses, we have identified quasi-periodic structures above 3\,$\sigma$, where $\sigma$ is the RMS of the Fourier intensity. For these pulses, we have estimated the substructure periods and widths. As mentioned above, the width estimates from the ACF might not always be reliable, but we report these for completeness. The estimated periods and widths of substructures are presented in Table~\ref{tab:bp}. The average width and period are 110(33)\,ms and 223(59)\,ms, respectively. The average substructure period is consistent with that from \cite{dong25_J1634}.
\par
For neutron stars, the microstructure quasi-period ($P_\mu$) is known to follow an empirical relationship with the rotation period ($P$). By fitting a power-law model, $P_\mu = A_p P^{\alpha_p}$, where $A_p$ is a scaling factor and $\alpha_p$ is the power-law index, we investigate whether LPTs also follow this relationship. The sample comprises pulsars, rotating radio transients (RRATs), magnetars, and LPTs. We adopt the spin period and geometric means and geometric standard deviations of $P_\mu$ for the pulsar, RRAT and magnetar populations from \cite{kld24}. The relevant measurements for LPTs are compiled form the literature, and are listed in Table~\ref{tab:lpt_microstructure}. For some of the LPTs, the $P_\mu$ measurement uncertainties are not available, and in such cases, we adopt a conservative 10\% multiplicative uncertainty. We perform the fit in logarithmic space and include the LPT measurements alongside the pulsar RRAT, and magnetar populations to investigate whether they follow the same empirical relation. The best-fitting relation is $P_\mu/(\mathrm{ms}) = (1.16 \pm 0.23)(P/(\mathrm{s}))^{(1.00 \pm 0.04)}$, which is shown in Figure~\ref{fig:microst}.

\begin{deluxetable*}{lrrcll}
\tablecaption{Rotation periods ($P$), geometric means of microstructure
quasi-periods ($P_\mu$), and geometric standard deviations (GSTD)
for known LPTs.}
\label{tab:lpt_microstructure}
\tablewidth{\textwidth}
\tablehead{
\colhead{Source} &
\colhead{$P$ (s)} &
\colhead{$P_\mu$ (ms)} &
\colhead{GSTD} &
\colhead{Reference} &
\colhead{Proposed nature}
}
\startdata
GLEAM-X J1627$-$5235 & 1091  & 6103  & 1.1 & \cite{kld24}      & Magnetar candidate$^{1}$ \\
GPM J1839$-$10       & 1318  & 1970  & 1.5 & \cite{men25}      & WD pulsar candidate$^{2}$ \\
GLEAM-X J0704$-$37   & 10497 & 12000 & 1.1 & \cite{hurley2024} & M-dwarf--WD binary$^{3}$ \\
CHIME/ILT J1634$+$44 & 841   & 216   & 1.3 & This work         & Unclassified \\
ASKAP J1839$-$0756   & 23222 & 2400  & 1.1 & \cite{lee25}      & NS candidate$^{4}$ \\
LPT J1555$-$5631     & 3732  & 24495 & 1.1 & \cite{hyman26}    & Unclassified \\
\enddata
\tablecomments{
$^{1}$ \citep{hurley2022};
$^{2}$ \citep{pelisoli25,horvath2026};
$^{3}$ \citep{rodriguez2025};
$^{4}$ \citep{lee25}.
}
\end{deluxetable*}

\subsection{Pulse Time of Arrival and Timing Analysis}\label{sec:timing}
Determining the arrival time of pulses with fine structures is challenging. One particular approach is to smooth the pulse profile to suppress the sub-pulse features and reveal the underlying average pulse envelope. Following \cite{dong25_J1634}, we convolve the pulse time series with a Gaussian kernel having a standard deviation of 200\,ms to smooth the pulse profile. Time at the peak of the smoothed profile is then taken as the time of arrival (ToA), and the full width at half maximum (FWHM) of the smoothed profile is considered as an estimate of the associated uncertainty on this ToA estimate. The chosen Gaussian smoothing scale is larger than the typical width of subpulse structures but smaller than the overall pulse duration, ensuring that the pulse envelope is retained. We note that for one particularly narrow pulse shown in the top left plot of Fig.~\ref{fig:pulse_gallery} (MJD 60938.57686), we use only a 6\,ms Gaussian kernel to smooth the profile. The estimated ToAs for all the detected pulses are presented in Table~\ref{tab:bp}. 
\par
\begin{figure*}
    \centering
    \includegraphics[width=0.6\linewidth]{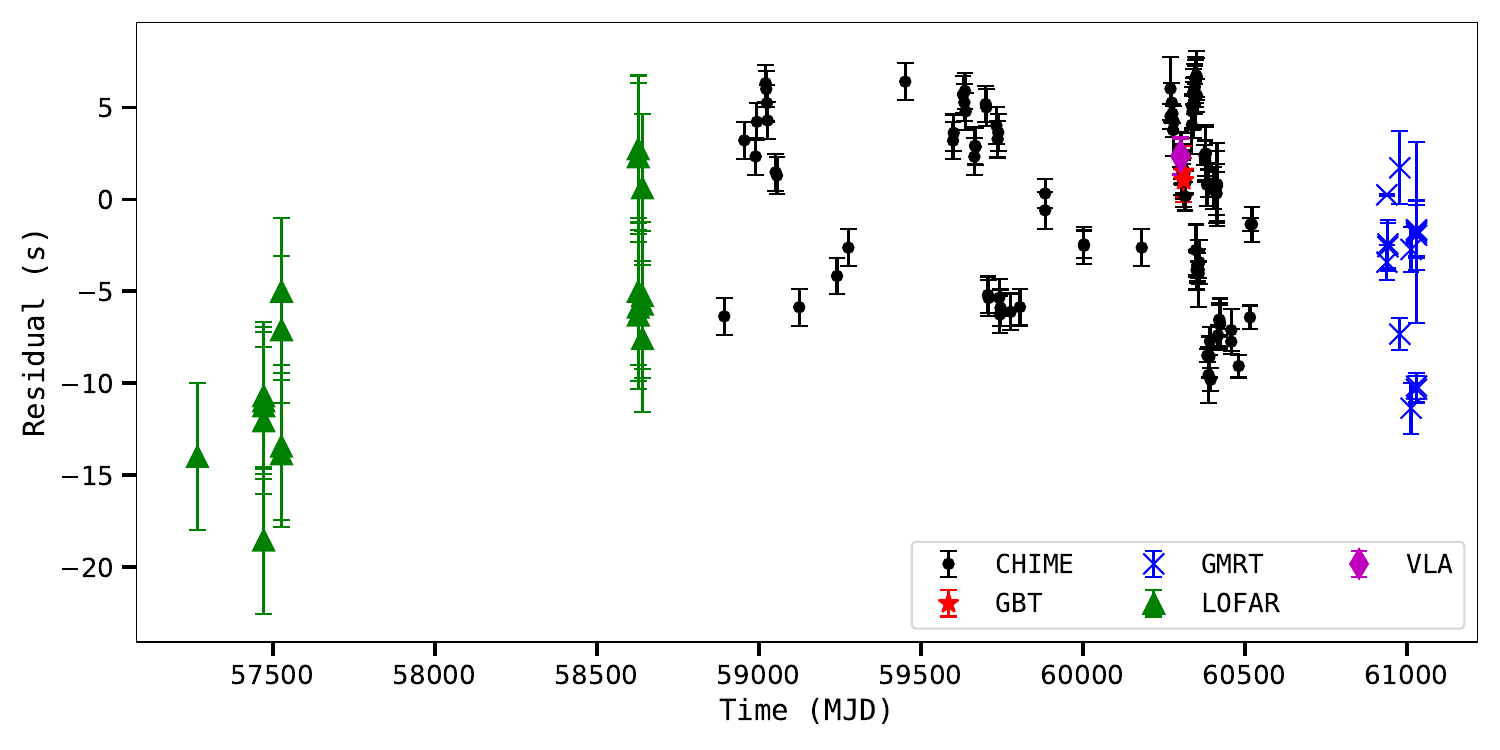}
    \caption{The figure shows timing residuals after fitting the spin frequency and its first time derivative to ToAs from our measurements as well as those from literature \citep{bvb25, dong25_J1634}.}
    \label{fig:timing}
\end{figure*}

Timing solution for J1634+44 were presented in the two independent discovery papers \citep{bvb25, dong25_J1634}. The individual data sets used in these papers have temporal baselines of roughly 4.5\,years (119 pulses, \citealt{dong25_J1634}) and 3.8\,years (19 pulses, \citealt{bvb25}). The combination of the arrival times of our pulses with these previously reported ToAs provides $\sim10.3$\,years long temporal baseline (152 pulses), which can help in better constraining the timing solution for this object. We used PINT\footnote{\url{https://github.com/nanograv/PINT}} \citep{luo2021} to estimate the timing solution. As mentioned in \cite{dong25_J1634}, ToAs for pulses detected using CHIME/FRB \citep{chime_frb2018} and CHIME/Pulsar \citep{chime_pulsar2021} instruments have an offset of 0.3(3)\,s, which is taken into account in our analysis. In addition, we searched for offsets between ToAs for pulses detected using CHIME and LOFAR, and CHIME and GMRT, by examining the post-fit residuals and reduced chi-squared values obtained with and without an additional free parameter, called ``{\tt jump}", between respective ToA sets. We do not find any significant {\tt jump} among the ToAs obtained from different observatories. Using all 152 ToAs spanning MJD~57268--61030 (referenced to MJD~59100), we fitted only for the spin frequency and its first time derivative. The timing model uses the DE442 Solar System ephemeris and the TT(BIPM2023) time standard. The best fitted spin period and its first derivative are $P=841.24684(2)\,\mathrm{s}$ and $\dot{P}=(-7.8\pm0.2)\times10^{-12}\,\mathrm{s\,s^{-1}}$, respectively, with a weighted RMS timing residuals of 2.5\,s. Fig.~\ref{fig:timing} shows the timing residuals with a weighted RMS of only 2.5\,s (i.e., a small fraction of the spin period) indicating a phase-connected timing model.

\section{Discussion}\label{sec:discussion}
\subsection{Radio Spectrum}\label{sec:disc_spectrum}
Spectral analysis is a useful tool for understanding emission physics and propagation effects. Here, we have conducted a detailed spectral study of J1634+44 using GMRT observations. Most of the single band spectral fits follow a power law. However, in a few cases a log-parabola fits better than a simple power law, for example, see the top right plot of Fig.~\ref{fig:spectral_fit_log}. In our sample, the simultaneous flux density measurements of four pulses at bands 3 and 4, and upper limits at band 5 frequencies suggest the radio spectra to be log-parabola (see Fig.~\ref{fig:spectral_fit_log}). The spectral turnover might be present in the other pulses too, but our single band detections have limited frequency coverage to sensitively probe such a possibility. We note that the only other LPT in which a spectral turnover has also been observed is ASKAP~J1839-0756 \citep{lee25}.
\par
The spectral turnover is well studied in NSs \citep[e.g.][]{kijak2011, jankowski2018, maan2022}. It can arise either intrinsically from the emission mechanism or due to absorption mechanisms like the free-free absorption and synchrotron self absorption. Assuming the spectral turnover to be due to free--free absorption, here we investigate the physical conditions of the absorber. Free-free absorption may be particularly relevant for LPTs in binary systems, where the emission is observed when the stellar wind from the companion interacts with the magnetic field of the primary compact object \citep[e.g.][]{horvath2026}. J1634+44 is also proposed to be in a binary system \citep{dong25_J1634, bvb25}, and \cite{bvb25} suggest that its emission may arise from magnetic interaction with the accreting material from the companion. The former possibility, i.e., magnetic interaction with the stellar wind, is also possible. In such cases, the spectral turnover and its variations can be naturally explained by the free-free absorption due to the accretion material or the stellar wind.
\par
\begin{figure*}
    \centering
    \includegraphics[width=0.497\linewidth]{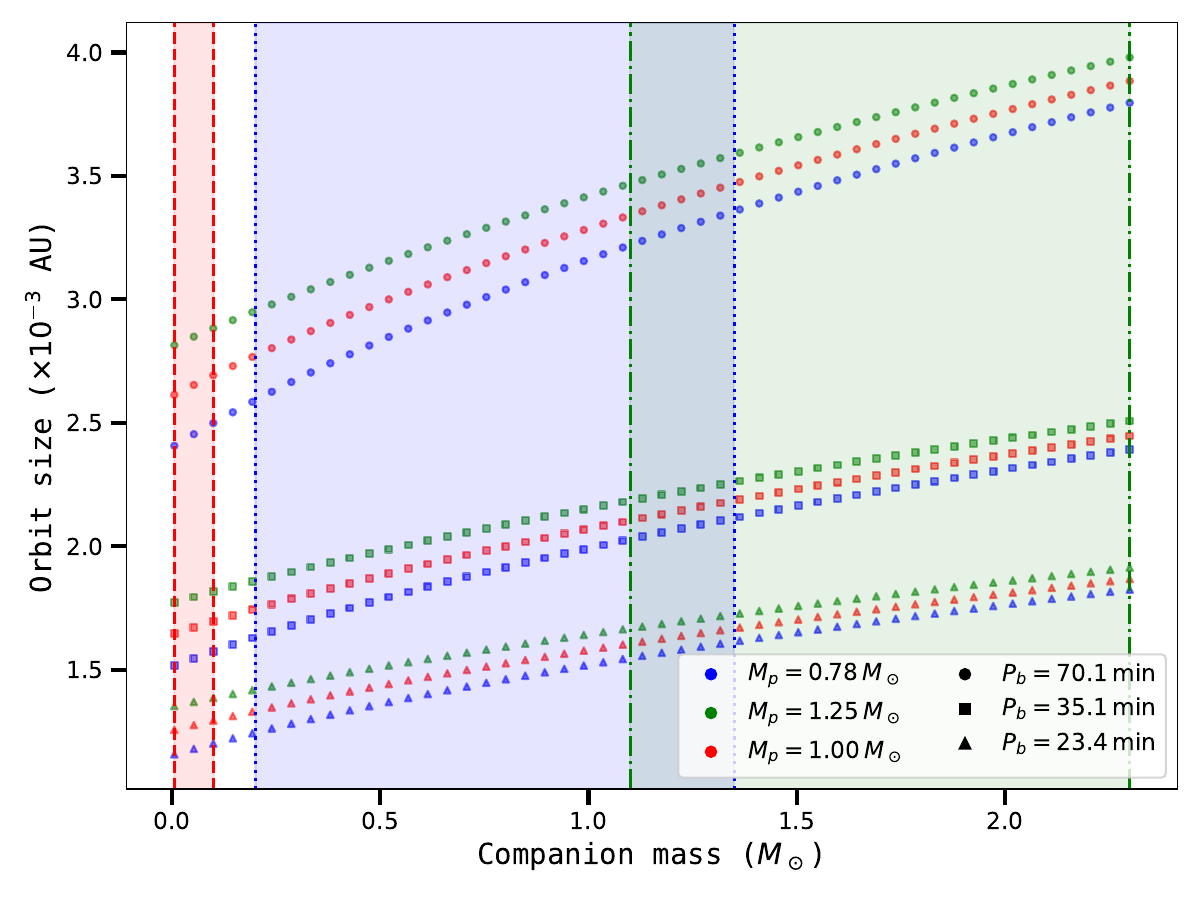}
    \includegraphics[width=0.497\linewidth]{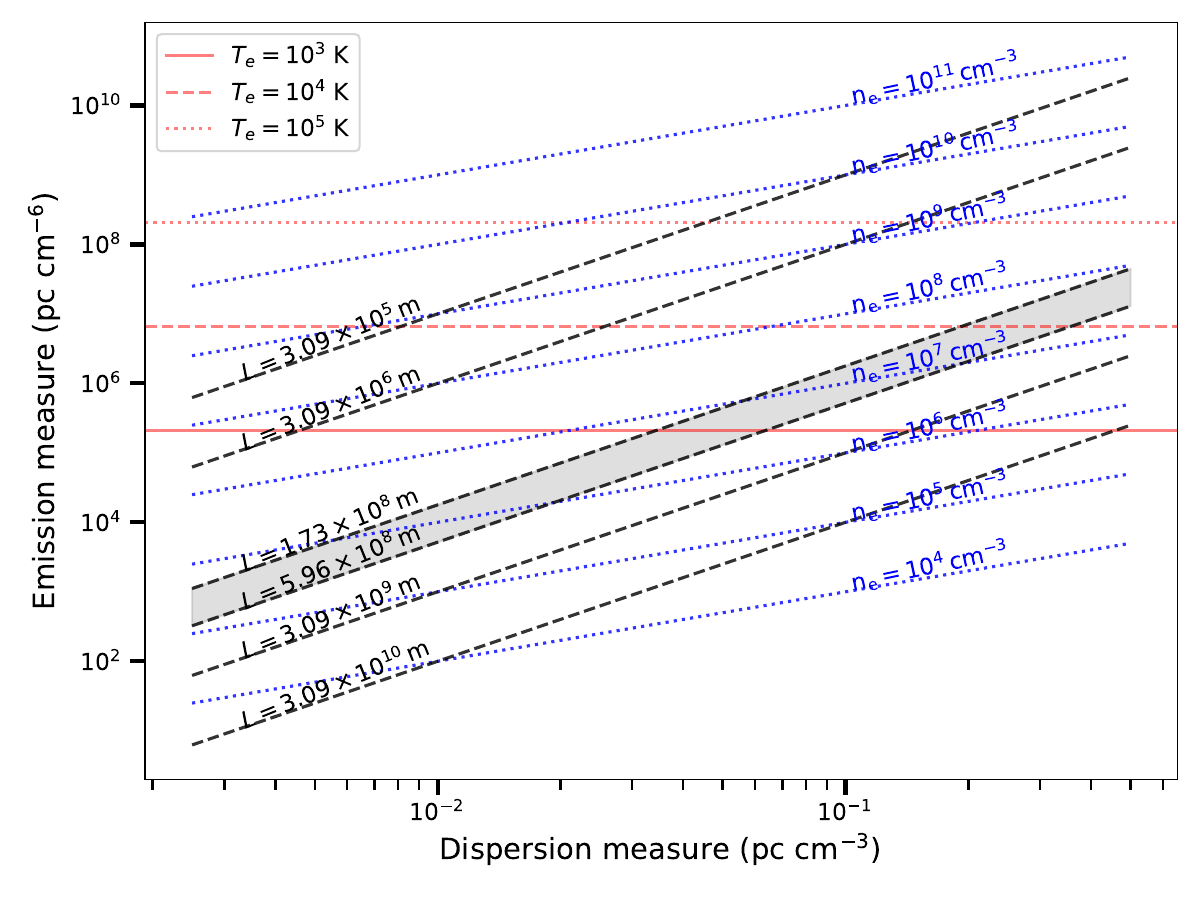}
    \caption{Left: the possible ranges of orbit size are shown for a range of companion masses, different primary masses ($M_P$), and orbital periods ($P_b$) corresponding to 5:1, 5:2, and 5:3 spin-orbit resonances. The different vertical shaded regions constrain the mass ranges for the companions, ultracool dwarf (0.005-0.1\,$M_\odot$), WD (0.2-1.35\,$M_\odot$), and NS (1.1-2.3\,$M_\odot$). Right: the EM-DM plane plot delineates constraints on the absorber. Inclined dotted and dashed lines represent the electron density and the thickness of the absorber, respectively, and the constant horizontal lines mark the electron gas temperatures. The possible orbital size of J1634+44 is represented in the grey shaded region that is between $(1.73-5.96)\times10^8$\,m. For more details, see the Section~\ref{sec:disc_spectrum}.}
    \label{fig:absorber_const}
\end{figure*}

We assume that absorption occurs within the orbit, and therefore, the absorber size is of the order of the orbital size. Further, we estimate the orbital size for all proposed combinations for this system. \cite{bvb25} have suggested that J1634+44 can be a WD binary system with the companion as another WD, a NS, or a UCD star. For binary systems, the orbital size can be estimated using Kepler's third law. We have considered a range of companion mass, 0.005-2.3\,$M_\odot$, which cover the possible mass range of UCD (0.005-0.1\,$M_\odot$), WD (0.2-1.35\,$M_\odot$), and NS (1.1-2.3\,$M_\odot$), and orbital period of 23.4 and 35.1\,minutes, corresponding to 5:3 and 5:2 spin-orbit resonances \citep{bvb25}. For the sake of completeness, we have also considered the orbital period of 70.1\,minutes corresponding to the 5:1 spin-orbit resonance. 
Based on the spectral energy distribution of J1634+44, \cite{bvb25} suggest that the WD's mass should be at least 0.78\,$M_\odot$. We have considered three values for the primary's mass, 0.78, 1.25, and 1\,$M_\odot$. Using the mentioned range for companion and primary's mass and orbital periods, we constrain that the orbital size should be in a range of $(1.2-4.0)\times10^{-3}$\,AU, i.e., $(1.73-5.96)\times10^8$\,m, as evident from the left panel of Fig.~\ref{fig:absorber_const}. 
\par
Furthermore, we constrain the electron density and temperature of the absorbing medium using the observed turnover frequency with the following relation \citep{rohlfs1996}:
\begin{equation}
\frac{\nu_t}{\mathrm{[GHz]}} = 0.1736
\frac{T_e^{-3/4}}{\mathrm{[K]}}
\sqrt{
\frac{\mathrm{EM}}{\mathrm{[pc\,cm^{-6}]}}
\left\langle g_{\rm ff} \right\rangle
},
\label{eq:turnover}
\end{equation}
where $\nu_t$ is the turnover frequency, $T_e$ is the electron gas temperature, EM is the emission measure, and $\langle g_{\rm ff}\rangle$ is the Gaunt factor, which is assumed to be unity. The emission measure is given by $\mathrm{EM}\approx n_e^2\delta d$, where $n_e$ and $\delta d$ are the electron density and thickness of the absorber, respectively. Assuming a uniform absorber, the DM contributed by the absorbing medium is $\delta \mathrm{DM}\approx n_e\delta d$. We consider a conservative range for $\delta$DM to be 0.01--2\% of the measured DM. This range is motivated by typical DM variations observed during eclipses of binary pulsars with low-mass companions, which are generally less than 1\% of their measured DM \citep[e.g.][]{stappers2001,bai2022}. In Fig.~\ref{fig:absorber_const}, we use the above-mentioned information to constrain the properties of the absorbing medium in the EM--DM plane. The observed turnover frequency defines the constant EM lines for different electron gas temperatures through Equation~\ref{eq:turnover}. We also show the corresponding electron-density relations, $\mathrm{EM} \approx n_e\mathrm{DM}$, together with the constraints for different absorber thicknesses, $\mathrm{EM} \approx \mathrm{DM}^2/\delta d$. The plausible range of absorber thickness is constrained by the possible orbital size highlighted by the grey-shaded region in the right-hand side panel. The confined region on the EM--DM plane that covers the shaded region constrains the allowed range of electron density and temperature of the absorbing medium. The allowed parameter space suggests an electron density of $\sim10^5$--$10^8\,\mathrm{cm^{-3}}$ and an electron gas temperature below $\sim10^5$\,K, depending on the DM contributed by the absorber.
\par
The inferred electron densities are substantially higher than those typical of the diffuse ISM but compatible with dense plasma environments expected in interacting binaries. A compact, dense absorber can produce significant free--free absorption while contributing only a modest DM. The free--free optical depth ($\tau_{\rm ff}$) scales approximately as $\tau_{\rm ff}\propto\int n_e^2\,dl$, whereas the DM depends linearly on the electron density. The DM of J1634+44 has remained stable since its discovery. This stability can therefore coexist with the gradual variations in the turnover frequency, which can result from changes in the density or geometry of a compact absorber and hence in its EM, without producing an appreciable change in DM. The allowed absorber temperature is below $\sim10^5$ K, corresponding to an extremely small thermal contribution at X-ray energies. Thus, the non-detection of X-ray emission from J1634+44 during its radio-active state \citep{ng2025ATel,lundy2025ATel} is consistent with a cool absorbing medium.
\par
The observed turnover frequencies show a gradual decrease, followed by a sudden jump at the last epoch (Fig.~\ref{fig:nu_p}). More systematic changes in turnover frequency has been observed for the magnetar XTE J1810-197 \citep{maan2022}. With the assumption of free-free absorption being the primary cause for the turnover, the changes can be explained by a decrease in the electron gas temperature or electron density or both. The sharp rise might indicate a stellar flare, which might increase the electron density and the gas temperature. 
\par
In our simultaneous multi-band observations, the pulses are not detected in band 5, i.e., at L-band (Fig.~\ref{fig:pulses_freq_evo}). However, we note that pulses from J1634+44 have been detected earlier at L-band and S-band \citep{dong25_J1634, shin25, Men2025ATel}. Given the relatively high flux density of the pulse detected at S-band \citep[0.3\,Jy,][]{Men2025ATel}, the most likely explanation for earlier detections at higher radio frequencies is potentially a shift of the turnover to adequately high frequency or highly dynamic nature of the individual pulse energies. The time difference between our first L-band observation and the VLA L-band observation \citep{shin25} was only ten days. If the VLA detection at L-band was indeed supported by a turnover in the spectrum at higher frequencies, the plasma conditions around the emission site are highly dynamic.

\subsection{Activity and Pulse Morphology}\label{sec:disc_pulse_mor}
In our observing campaign, the first pulse from J1634+44 was detected on 20 September 2025. It was the first observation of our campaign, and nine days before the first pulse report by \cite{shin25}. This pulse has a single peak profile, belonging to C1 morphology, with the lowest detected fluence (13.3\,Jy\,ms) in our sample. Subsequently, after two rotations of a period of 841\,s, another pulse was detected with three subpulses in ascending order of the peak flux density and with C2 morphology. The non-detection of J1634+44 at CHIME before 29 September 2025, and detection of first a narrow and weak pulse followed by a more feature-rich, triple-peaked pulse in our campaign indicate that the source likely transitioned into an active state around the time of our first observation. Based on this scenario and assuming that the pulse reported by \citet{dong25_J1634} was the last detection before J1634+44 entered into a quiescent state, it took $\sim$1.14\,years before its reactivation. 
\par
In our sample, the majority of pulses fall into the C4 category, with multi-peaked emission reaching the baseline in between the peaks. Such pulse variations have been seen in pulsars \citep[e.g.][]{cordes1979}, magnetars \citep[e.g.][]{camilo2006, maan2019}, and FRBs \citep[e.g.][]{hessels2019, cordes2019}. The morphological features are likely related to the emission mechanism, potentially further modulated by the propagation-related effects. Multiple pulse components might arise from a structured radio beam, analogous to the core–cone or patchy-beam configurations proposed for pulsars \citep[e.g.][]{rankin83, lyne88, mitra1999_pulsar_beam}. However, given the sporadic nature of LPT pulses, the emission regions might be highly dynamic, with variations in plasma conditions or magnetic-field configuration potentially producing changes in pulse morphology, similar to those in magnetars \citep[e.g.][]{kaspi2017}. Alternatively, if the LPT emission originates from magnetized white-dwarf binaries, interactions between the stellar magnetospheres or with the companion winds can generate localized particle-acceleration regions. In such systems, changes in magnetic-field strength and geometry or in the location of the interaction region, could naturally produce spectro-temporal variations in the emitted radiations.
\par
In one of the detected pulses, we have identified very narrow spiky features, as shown in bottom left plot of Fig.~\ref{fig:pulse_gallery} (around 1.5\,s). Such features are also seen in DART J1832-0911 on top of the broad pulse \citep{li2024_dartJ1832-0911}. However, in J1634+44, these narrow pulses are slightly off from the main pulse component. Notably, these features become very faint, and some of them are not detectable at band 4. The frequency-dependent pulse variations and the steep spectrum at band 4 can explain the faintness and non-detection there. Such features might result from the activation of smaller emission regions in the vicinity of the main pulse emission region. 

\subsubsection{Quasi-periodicity}
Quasi-periodic microstructures have been observed in all variants of neutron stars \citep[e.g., see][for an overview]{kld24} and a few FRBs \citep[e.g.,][]{pastor2023}, and they carry information about the underlying emission physics \citep[e.g.,][]{cordes1979, Mitra_2015}. Multiple hypotheses have been proposed to explain the origin of such microstructures. Two widely considered possibilities invoke either angular beaming or temporal modulation of the emission \citep[e.g.,][]{cordes1979, Mitra_2015}. In the former scenario, the quasi-periodic microstructures arise as the line of sight successively samples multiple discrete emission regions, whereas in the latter, they reflect intrinsic temporal modulation of the emission. 
\par
In four of our detected pulses, we have seen clear microstructures, reminiscent of those seen from NSs (Fig.~\ref{fig:microst}). Observationally, for neutron stars, the period of quasi-periodic microstructures has been found to increase linearly with the rotation period \citep[e.g.][]{cordes1979, kramer02}. The mean quasi-period of microstructures in J1634+44 seems broadly consistent with the prediction of the $P$–$P_\mu$ relation by \citet{kld24} but differ significantly from that reported by \citet{amarnath26}. We note that \citet{amarnath26} did not use the 76 second pulsar in their $P$–$P_\mu$ fit which might have resulted in a slightly different slope. 
\par
We extended the $P$–$P_\mu$ analysis to include LPTs. The best-fit power law yields an index of $\alpha_p = 1.00 \pm 0.04$ and a scaling factor of $A_p = 1.16 \pm 0.23$, indicating an approximately linear relationship between the rotation period and microstructure quasi-period. Interestingly, both the confirmed WD–M-dwarf binary GLEAM-X J0704$-$37 and the proposed WD-pulsar candidate GPM J1839$-$10 closely follow this relation (Fig.~\ref{fig:microst}). This observation leads to an important deduction: the $P$–$P_\mu$ relation is not just limited to neutron stars but might be extendable to coherent radio emission from other compact objects like white dwarfs. The measurements for LPTs J1634+44 and J1555$-$5631 deviate slightly from the above fitted relation. Two LPTs proposed to be magnetar or NS candidates, GLEAM-X J1627$-$5235 and ASKAP J1839$-$0756, exhibit significant deviations (Fig.~\ref{fig:microst}). We note that only single or limited $P_\mu$ measurements are available for four LPTs and we have assumed a 10\% uncertainties on those. It is possible that further measurements of $P_\mu$ and robust estimates of the associated uncertainties would make the above discussed four LPTs too more consistent with the $P$–$P_\mu$ relationship. Alternatively, the deviations might reflect intrinsic differences in emission mechanism or those in magnetospheric conditions. A larger, more homogeneous sample of $P_\mu$ measurements for LPTs is needed to assess the applicability of the $P$–$P_\mu$ relationship to the whole LPT population. Nevertheless, GPM~1839$-$10 and GLEAM-X~J0704$-$37 closely following the relationship strongly argue that the $P$–$P_\mu$ relationship is not limited only to the neutron stars but potentially followed by coherent emission from compact sources in general.

\subsection{Pulse Time of Arrival Analysis}
\begin{figure*}
    \centering
    \includegraphics[width=0.497\linewidth]{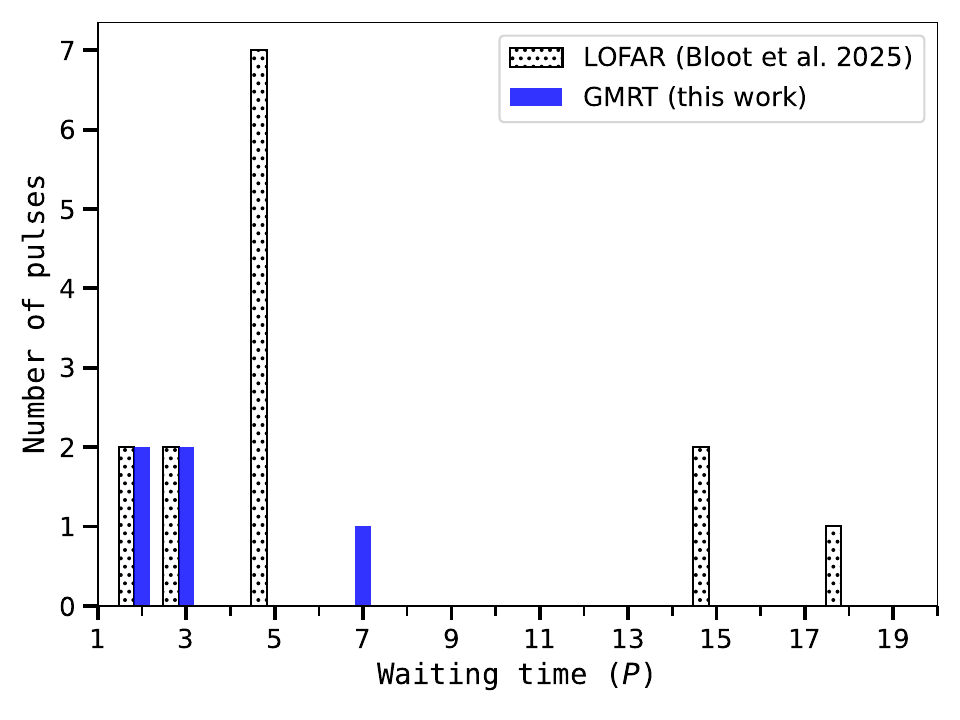}
    \includegraphics[width=0.497\linewidth]{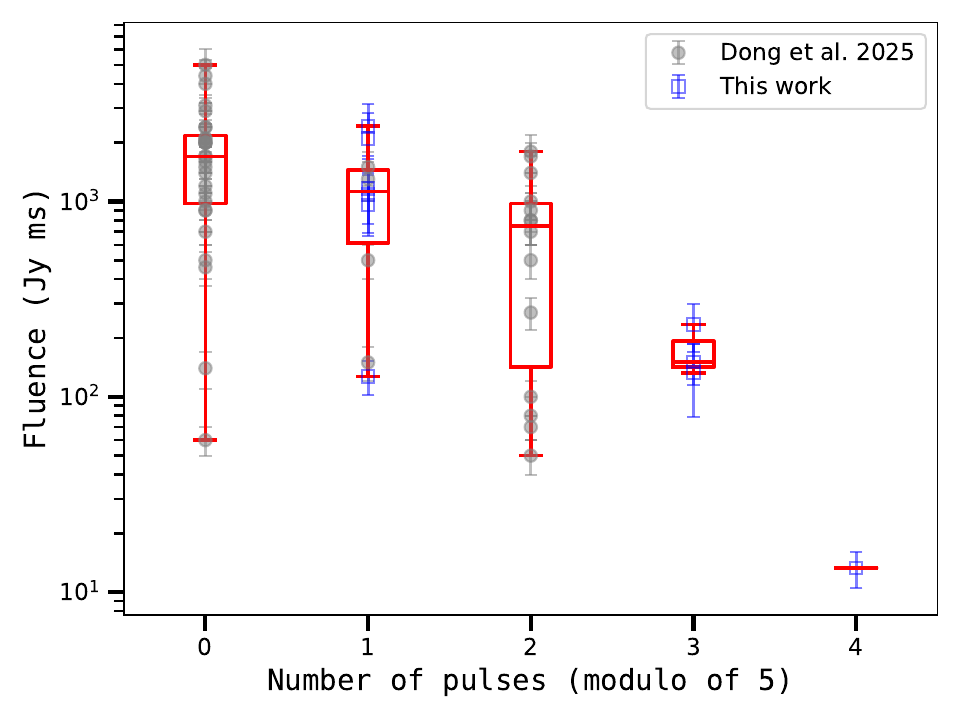}
    \caption{Left: the distribution of pulse waiting times. LOFAR \citep{bvb25} and GMRT data show a preference for two or three pulses in every five rotations of 841.2468\,s period. Right: the fluence variation with number of rotations mod 5, for pulses detected by CHIME/GBT/VLA \citep{dong25_J1634} and in this work. The x-axis represents mod 5 of the number of rotations from the first detected pulse by \citet{dong25_J1634}.}
    \label{fig:wait_time}
\end{figure*}

In our sample, typically two pulses have been detected every five rotations, with the rotation period of 841.2468\,s. As shown in Fig.~\ref{fig:wait_time}, the waiting times of two or three rotations are favored over the other possibilities. A similar pattern was observed in LOFAR data, leading \citet{bvb25} to propose a 5:2 or 5:3 spin-orbit resonance model, which is supported by our GMRT data. In the spin-orbit resonance scenario, one might expect a statistical trend in the fluence estimates over different rotations. Fig.~\ref{fig:wait_time} shows fluence values of our sample as well as those reported by \cite{dong25_J1634} with the absolute number of rotations from the first pulse detected by CHIME \citep[MJD~58893.62626,][]{dong25_J1634} mod 5. The overall sample shows a significant parabolic trend, further confirming the presence of a secondary period that is five times the shorter period. Strict and stable 5:2 and 5:3 resonances would have resulted into bimodal and trimodal distributions in Fig.~\ref{fig:wait_time}. There is no strong evidence for a bimodal distribution. There seems to be a weak evidence for a 5:3 resonance as the numbers and fluences at the last two bins in the right plot of Fig.~\ref{fig:wait_time} are significantly less than those at the others. The observed fluence pattern could also result if the above possible resonances are not stable or strict. Finally, having the secondary period slightly different from strictly 5 times the primary period as the orbital period could also result in the observed pattern of fluences shown in Fig.~\ref{fig:wait_time}.
\par
The combination of all ToAs, from previous works \citep{bvb25, dong25_J1634} and this work, has enabled more precise timing of J1634+44. The refined estimates for $P$ and $\dot{P}$ values are 841.246835(3)\,s and $\rm -7.79(3)\times 10^{-12}\,s\,s^{-1}$, respectively. These new estimates of the period and its derivative are $\sim55\%$ and $\sim75\%$, respectively, more precise than those reported by \cite{dong25_J1634}. J1634+44 has been proposed to be in the binary system \citep{dong25_J1634, bvb25}, however, its binary model is entirely unconstrained. \cite{bvb25} have proposed a 5:2 or 5:3 spin-orbit resonance model, which suggests an orbital period of 2103\,s and 1402\,s, respectively. The longer period 4206\,s can also be the orbital period. Given the longer temporal baseline of around a decade, we have attempted to fit a binary model to the ToAs using these three orbital periods and examined the fit quality based on the reduced chi-squared and residuals. In all cases, the fit either improved marginally or did not converge (see more details in Appendix~\ref{sec:resonant_orbits}). Therefore, based solely on the timing analysis of 152 ToAs over a decade-long temporal baseline, we cannot constrain the orbital period to correspond to one of the proposed spin-orbit resonance models. Nevertheless, as discussed above, the waiting times of the pulses detected by LOFAR and GMRT still favor a spin-orbit resonance.

\section{Conclusion}\label{sec:conclusion}
We have presented a spectro-temporal and spectral study of the LPT~J1634+44 using single-band and simultaneous multi-band GMRT observations. The source exhibits pronounced pulse-to-pulse variability: the first detected pulse has a single-peaked profile, whereas pulses at later epochs show complex substructures, and the fluence varies by roughly two orders of magnitude (13--2460\,Jy\,ms) over the short span of our campaign. Our simultaneous observations spanning 300--1460\,MHz reveal significant frequency-dependent subpulse evolution and interesting spectra. The pulse spectra are variable and exhibit a clear turnover, with the turnover frequency decreasing systematically over the epochs, except in one of the last observations, where it varies substantially from pulse to pulse. Interpreting the turnover as free-free absorption, we constrain the properties of the absorbing medium and show it to be consistent with the intra-binary material. We also attempted to model the ToAs with binary models motivated by the spin-orbit resonance configurations proposed by \cite{bvb25}, but these fits yield only marginal improvements or fail to converge and remain inconclusive. Nevertheless, the waiting time between the successive pulses and the fluence variations present a weak evidence supporting a 5:3 resonance or an orbital period 5 times the primary period. Finally, using our measurements of quasi-periodic microstructures of J1634+44 and those of other LPTs from the literature, we demonstrate that the $P$--$P_\mu$ relationship is not limited to only neutron stars but is also followed by potentially other compact objects emitting coherent radio signals, including some LPTs identified as white dwarfs.  

\begin{acknowledgments}
AK thanks Jayanta Roy for sharing the script for reading the GMRT full-polar raw data. We thank the GMRT staff for making these observations possible. The GMRT is run by the National Centre for Radio Astrophysics (NCRA) of the Tata Institute of Fundamental Research (TIFR). This work is supported by the Department of Atomic Energy (DAE), Government of India, under the project ``Next Generation Instrumentation for Radio Astronomy'' (Project UID: RTI4017).
\end{acknowledgments}

\facility{GMRT}

\software{astropy \citep{astropy18, astropy22},  
          numpy \citep{numpy20},
          matplotlib \citep{matplotlib07},
          RFIClean \citep{MvLV21},
          DSPSR \citep{vsb11},
          PRESTO \citep{ransom11},
          PINT \citep{luo2021}, and
          TEMPO2 \citep{edwards_tempo2_2006}.
          }

\appendix
\section{Testing Resonant Binary Models}
\label{sec:resonant_orbits}
In Section~\ref{sec:timing}, we presented pulse timing analysis assuming J1634+44 to be an isolated system. The best-fit timing model yields a weighted root-mean-square (WRMS) of 2.5\,s and a reduced $\chi^2$ of 29.9. Even excluding ToAs with error greater than 2\,s only marginally improves the timing fit. It indicates that the higher post-fit residuals are either intrinsic timing noise or unmodeled orbital variations, if any. Here, we test the proposed spin-orbit resonance scenarios \citep{bvb25} using the pulse timing analysis. The DD86 binary model \citep{damour_general_1986} is used to account for potential relativistic dynamics, which is implemented in \texttt{tempo2} \citep{edwards_tempo2_2006}. We fit for spin parameters ($F0, F1$) and Keplerian orbital parameters (orbital period $P_B$, time of passage through periastron $T_0$ and projection of semi-major axis $A_1$), while initially freezing the longitude of periastron ($\omega$) and eccentricity ($e$) to be zero. The orbital geometry of this source is entirely unconstrained, and freezing $\omega$ and $e$ prevents non-convergence of the fit. This approach has allowed us to evaluate whether the proposed resonant orbital periods could absorb the timing residuals, independent of specific geometric orientations. The outcomes of these trials are summarized in Table~\ref{tab:orbits}.
\begin{table}[h]
    \centering
    \begin{tabular}{l c c c c}
         \hline\hline
         Model &  $P_B$ & WRMS & Reduced $\chi^2$ & Mass Function \\
         & (s) & (s) & & ($M_\odot$) \\
         \hline
         \multicolumn{5}{c}{\textit{$e=0$, $\omega=0$}} \\
         \hline
         5:1 & 4206.26 & 2.29 & 26.05 & 9.42 \\
         5:2 & 2103.14 &  2.06 & 21.13 & -72.89 (unphysical) \\
         5:3 & 1402.12 &  1.35 & 9.07 & -575.54 (unphysical)\\
         \hline
    \end{tabular}
    \caption{Parameters derived from trial binary model fits using resonant orbital periods. We initialize the fitting by assuming that the eccentricity ($e$) and the longitude of periastron ($\omega$) are zero. Adding binary model free parameters to the fitting reduces the residuals; however, in the 5:2 and 5:3 spin-orbit resonance scenarios, the resulting mass functions become unrealistic.}
    \label{tab:orbits}
\end{table}
\par
Although introducing additional orbital parameters reduces WRMS and the reduced $\chi^2$ values compared to the isolated model, the resulting best-fit and derived parameters become unrealistic. In particular, the projected semi-major axis becomes negative, resulting in unphysical mass functions. Since these unconstrained fits yielded absurd mass functions and geometrically impossible negative eccentricities, a robust comment supporting the spin-orbit resonance model cannot be made with the current dataset.

\bibliography{ref}{}
\bibliographystyle{aasjournalv7}

\end{document}